\documentclass[floats,floatfix,showpacs,amssymb,prd,twocolumn,superscriptaddress,nofootinbib,reprint]{revtex4-2}

\usepackage{amssymb,amsmath,verbatim,mathtools,needspace,enumitem,etoolbox,graphicx,physics,microtype,afterpage,bigints,textcomp,gensymb,tabularx,lmodern}
\usepackage[dvipsnames,table, usenames]{xcolor}
\definecolor{lightgray}{gray}{0.9}
\definecolor{lightblue}{rgb}{0.93,0.95,1.0}
\definecolor{linkcolor}{rgb}{0.846416, 0.272473, 0.421631}
\usepackage[unicode, colorlinks=true, linkcolor=linkcolor, citecolor=linkcolor, filecolor=linkcolor,urlcolor=linkcolor, pdfusetitle]{hyperref}
\usepackage[all]{hypcap}
\usepackage[T1]{fontenc}
\usepackage[utf8]{inputenc}
\usepackage{siunitx}
\usepackage{enumitem}
\usepackage[normalem]{ulem}
\usepackage{bm}
\usepackage{booktabs}

\DeclareSIUnit\year{yr}

\newcommand{\sub}[1]{_{\text{#1}}}
\newcommand{\super}[1]{^{\text{#1}}}

\newcommand\orcid[1]{\href{https://orcid.org/#1}{$\;\!$\includegraphics[width=0.7em]{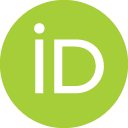}}}

\newcommand{\bham}{
\affiliation{Institute for Gravitational Wave Astronomy
\& School of Physics and Astronomy, University of
Birmingham, Birmingham, B15 2TT, United Kingdom}}

\def\aj{\rm{AJ}}

\def\apjl{\rm{ApJ}}

\def\aap{\rm{A\&A}}

\def\mnras{\rm{MNRAS}}
\def\prd{\rm{PRD}}
\def\prl{\rm{PRL}}
\def\prx{\rm{PRX}}

\def\nat{\rm{Nature}}

\def\cqg{\rm{CQG}}
\def\lrr{\rm{LRR}}

\newcommand{\ssim}{\mathchar"5218\relax\,}

\newcommand\ldc{LDC--1}
\newcommand\source{SmBBH}

\hypersetup{pdfauthor={Name}}

\begin{document}

\title{Bayesian parameter estimation of stellar-mass black-hole binaries with LISA}

\author{Riccardo Buscicchio\orcid{0000-0002-7387-6754}}
\email{riccardo@star.sr.bham.ac.uk}
\bham

\author{Antoine Klein\orcid{0000-0001-5438-9152}}
\bham

\author{Elinore Roebber\orcid{0000-0002-5709-4840}}
\bham

\author{Christopher~J.~Moore\orcid{0000-0002-2527-0213}}
\bham

\author{Davide Gerosa\orcid{0000-0002-0933-3579}}
\bham

\author{Eliot  Finch\orcid{0000-0002-1993-4263}}
\bham

\author{Alberto Vecchio\orcid{0000-0002-6254-1617}}
\bham

\date{\today}

%%%
\begin{abstract}
We present a Bayesian parameter-estimation pipeline to measure the properties of inspiralling stellar-mass black hole binaries with LISA.
Our strategy (i) is based on the coherent analysis of the three noise-orthogonal LISA data streams, (ii) employs accurate and computationally efficient post-Newtonian waveforms ---accounting for both spin-precession and orbital eccentricity--- and (iii) relies on a nested sampling algorithm for the computation of model evidences and posterior probability density functions of the full 17 parameters describing a binary.
We demonstrate the performance of this approach by analyzing the LISA Data Challenge (\ldc{})
dataset, consisting of 66 quasicircular, spin-aligned binaries with signal-to-noise ratios ranging from 3 to 14 and times to merger ranging from 3000 to 2 years.
We recover 22 binaries with signal-to-noise ratio higher than 8. Their chirp masses are typically measured to better than $0.02 M_\odot$ at 90\% confidence, while the sky-location accuracy ranges from 1 to 100~square degrees.
The mass ratio and the spin parameters can only be constrained for sources that merge during the mission lifetime.
In addition, we report on the successful
recovery of an eccentric, spin-precessing source at signal-to-noise ratio 15 for which we can measure an eccentricity of $3\times 10^{-3}$ and the time to merger to within $\sim 1$~hour.
\end{abstract}
%%%

%%%
\maketitle
%%%

%%%
\section{Introduction}
\label{sec:intro}

%%%
The Laser Interferometer Space Antenna (LISA)~\cite{2017arXiv170200786A} is a gravitational-wave (GW) observatory targeted at the discovery and precise study of compact binary systems ranging from white dwarfs of masses $\ssim 0.1$--$1\,M_\odot$ to black holes with masses up to $\ssim 10^7\,M_\odot$.
Cosmological phenomena with characteristic timescale between $\ssim 1\,\mathrm{hr}$ and $\ssim 10\,\mathrm{sec}$ might also be detectable.

One of the sources of great interest are stellar-mass binary black holes
(hereafter \source{}s, also referred to as  stellar-origin binary black holes, SOBBHs\footnote{We prefer to use SmBBHs instead of SOBBHs to acknowledge that the problem of detecting and characterizing these sources is independent of the (astro)physics that determines their formation.
})
in the mass range $\ssim 10$--$100\,M_\odot$ which populate LISA's sensitivity window at frequencies $f \gtrsim 10\,\mathrm{mHz}$. These systems are now routinely observed merging at $\sim 100\,\mathrm{Hz}$
by the ground-based laser interferometers LIGO and Virgo \cite{2019PhRvX...9c1040A,2021PhRvX..11b1053A}.
LISA is expected to observe  \mbox{$\ssim 1$--$10$} \source{}s during the whole mission~\cite{2016PhRvL.116w1102S, 2016MNRAS.462.2177K,2018PhRvL.121y1102W,2019MNRAS.488L..94M}, an estimate that crucially depends on the upper-mass cutoff of \source{}s, the detection strategy, as well as the LISA performance at high frequencies. Each of these systems will contain valuable information in terms of both astrophysical formation channels \cite{2016PhRvD..94f4020N,2016ApJ...830L..18B, 2018MNRAS.481.5445S,2019PhRvD..99j3004G} and fundamental physics constraints~\cite{2017PhRvD..96h4039C,2019PhRvD.100f4024G,2019PhRvD..99l4043T,2020PhRvD.101j4038T}.  The subset of sources that merge on a timescale of $O(10)$ yr will be even more unique, allowing for its multiband characterization using GWs from both space and the ground \cite{2016PhRvL.117e1102V}, as well as advanced planning of electromagnetic follow-up campaigns~\cite{2016PhRvL.116w1102S, 2020NatAs...4...26M}.

Because of the long duration and complex morphology of their signals, detecting and characterizing the physics of \source{}s with LISA is a highly nontrivial challenge.
Previous work has been carried to study parameter estimation for circular precessing systems~\cite{2020PhRvD.102l4037T}.
In this context, we tackle for the first time the effects of orbital eccentricity coupled with spin precession.

\begin{figure}[tbp] 
    \centering
    \includegraphics[width=\columnwidth]{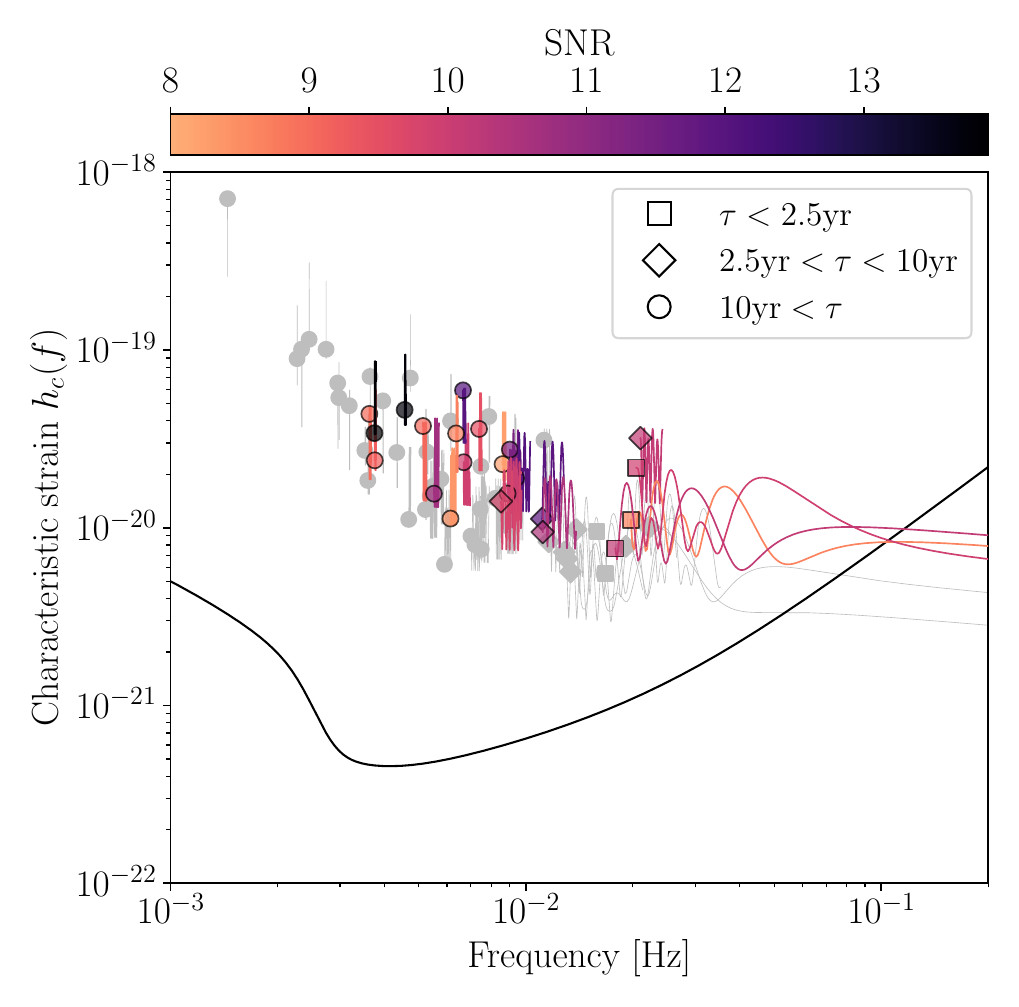}
    \caption{Characteristic strain of the injected challenge dataset sources. For low-frequency, quasimonochromatic sources, the characteristic strain is modulated by the LISA orbit throughout the mission.
    Markers denote the sources' initial frequencies, continuous lines denote their spectral characteristic strain amplitude. Binaries merging within the dataset duration of $\SI{2.5}{\year}$ are marked by squares, while diamonds and circles indicate binaries coalescing in $2.5$-$10$ and over 10 years, respectively. More details on the source properties are outlined in Sec.~\ref{sec:analysis}.
    Lines and markers are colored according to the signal-to-noise ratios (SNRs);
    unresolved sources with $\text{SNR}<8$ are grayed. Note that characteristic strain amplitudes were constructed using the low-frequency approximation of the LISA response~\cite{1998PhRvD..57.7089C}.
    The solid black line denotes the LISA characteristic noise spectral amplitude~\cite{2019CQGra..36j5011R, 2015CQGra..32a5014M}.
    }
    \label{fig:Catalog}
\end{figure}

The importance of \source{}s for the LISA science case is well recognized. To this end, a set of LISA data challenges (LDCs) are in progress under the auspices of the LISA consortium as part of the core  preparation activities for the mission adoption (see  \href{https://lisa-ldc.lal.in2p3.fr/}{lisa-ldc.lal.in2p3.fr}).
The first set of these challenges (\ldc{}) contains mock datasets populated with
\source{}s. 
The analyzed systems from \ldc{} are illustrated in Fig.~\ref{fig:Catalog}. 
Most of the sources appear as quasimonochromatic.
However, a few of them merge within the observing time (which in \ldc{} was set to $2.5$ yr), thus allowing a finer characterization of their parameters through the chirping morphology. 

Here we report on the results of the
analysis of all \source{}s in \ldc{} using the generic Bayesian codebase we are developing, hereafter referred to as \textsc{Balrog}. While the \ldc{} sources  were injected assuming quasicircular binaries with aligned spins, we also present preliminary results on the more general problem of analyzing systems with orbital eccentricity and spin precession.
Overall, this paper quantifies how well \source{}s can be characterized with LISA once they have been detected.

This paper is organized as follows. In  Sec.~\ref{sec:analysis}, we describe our data analysis strategy and outline its technical implementation.
In Sec.~\ref{sec:results}, we present the challenge dataset we analyzed and our inference results.
In Sec.~\ref{sec:concl}, we provide our conclusions and pointers to future work.
Throughout the paper we use $G=c=1$.
%%%
\section{Analysis approach}
\label{sec:analysis}
%%%

\subsection{Inspiralling stellar-mass black-hole binaries}

\source{}s in the early inspiral region probed by LISA are long-lived sources radiating for most or all of the mission duration, depending on their masses and orbital period at the start of the mission. In fact, for a binary with component masses $m_1$ and $m_2$ at redshift $z$, the leading Newtonian order time to coalescence is \cite{1963PhRv..131..435P,2014LRR....17....2B}
\begin{equation}
\label{eq:mergertime}
    \tau \approx 4.1\!\left(\frac{\nu}{1/4}\right)^{-1}\!\left(\frac{f}{20\,\mathrm{mHz}}\right)^{-8/3}\!\left(\frac{M_z}{50\,M_\odot}\right)^{-5/3}\!\mathrm{yr}\,,
\end{equation}
where $M_z = (1+z)(m_1 + m_2)$ is the redshifted total mass,  $\nu = m_1 m_2/(m_1+m_2)^2$ is the symmetric mass ratio, and $f$ is the GW frequency.
During this period the (leading Newtonian order) number of wave cycles to merger is
\begin{equation}
    {\cal N}\!\approx\!4.1\!\times\!10^6\! \left(\frac{\nu}{1/4}\right)^{-1}\!\!\!\left(\frac{f}{20\,\mathrm{mHz}}\right)^{-5/3}\!\!\left(\frac{M_z}{50\,M_\odot}\right)^{\!-5/3}\!.
\end{equation}
Consequently, if the source merges in a few years, i.e. unless $\dot{f}T_\mathrm{obs}\!\ll\!f$, most of the wave cycles are accumulated in the LISA band.
These cycles need to be matched by the analysis, in sharp contrast with the current LIGO--Virgo observations, for which only a few or tens of cycles are in band.

The signal will also have complex features induced by spin-precession and the effects of eccentricity.
This adds complexity to the waveform and to the structure of the likelihood function, and increases the dimensionality of the parameter space.

If the black-hole spins are misaligned with the orbital angular momentum, this will induce a precession of the orbital plane around the axis of the total angular momentum characterized by a number of spin-precession cycles before merger~\cite{1994PhRvD..49.6274A,2021arXiv210610291K}
\begin{align}
    {\cal N}_\mathrm{spin} &\approx 1.9\times 10^3 \left(1 - \frac{\delta\mu^2}{7}\right)\, \left(\frac{\nu}{1/4}\right)^{-1}
    \\
    &\times\left(\frac{f}{20\,\mathrm{mHz}}\right)^{-1} \left(\frac{M_z}{50\,M_\odot}\right)^{-1}\,, \nonumber
\end{align}
where $\delta\mu = (m_1 - m_2)/(m_1 + m_2)$ is the dimensionless mass difference.

Eccentricity may also be non-negligible in the LISA band, and an important parameter to measure as it is a tracer of the environment in which these binaries reside and the formation channel(s) of these systems.
The number of periastron precession cycles before merger is~\cite{1963PhRv..131..435P, 2004PhRvD..70f4028D}
\begin{equation}
    {\cal N}_\mathrm{ecc}\!\approx\!6.4\times\!10^3\!\left(\frac{\nu}{1/4}\right)^{-1}\!\!\left(\frac{f}{20\,\mathrm{mHz}}\right)^{-1}\!\!\left(\frac{M_z}{50\,M_\odot}\right)^{-1}.
\end{equation}
Note that these estimates are valid in the low-eccentricity limit.

It is therefore clear that to accurately reconstruct the physics of \source{}s, one needs to deal with the full complexity of the 17 dimensions parameter space that describes GWs radiated by a binary system in general relativity.
The morphology of these \source{} signals is very different from both those currently observed by LIGO and Virgo as well as the supermassive BBH merger signals expected in LISA. In fact, these signals have more in common with the extreme-mass-ratio inspiral (EMRI) signals also expected in LISA which also contain $10^5 - 10^6$ wave cycles in band and can exhibit strong relativistic precession effects. The data analysis challenges presented by this source type are well-known \cite{2011CQGra..28i4016C, 2017PhRvD..96d4005C, 2017PhRvD..95j3012B}. In addition, EMRI present a severe modelling challenge, see e.g. \cite{2020PhRvL.124b1101P, 2021PhRvL.126e1102C, 2018CQGra..35n4003V}.

\subsection{Statistical inference}

In this paper we are not concerned with the (significant) challenge of actually searching for \source{}s~\cite{2019MNRAS.488L..94M}, but we restrict ourselves to the problem of measuring the source parameters once candidates have been initially identified through a first search stage. We will therefore assume that a preceding pipeline provides an initial, possibly poor guess of the source parameters on which we can deploy our Bayesian parameter-estimation approach.

Our analysis is performed using the three noise-orthogonal time-delay-interferometry (TDI) outputs that are generated by combining the readouts of the LISA phase-meters~\cite{2002PhRvD..66l2002P}. This stage suppresses by a factor $\approx 10^8$ the laser noise leaving the data stream only affected by the secondary noise sources and GWs. The details of this complex procedure are under active investigation and development, see e.g. Refs.~\cite{2019PhRvD..99h4023B,2021PhRvD.103h2001V,2021PhRvD.104b3006B,2021PhRvD.103l3027H}.

We employ a coherent analysis of the full LISA TDI outputs, $d = \{d_k; k = A,E,T\}$,
by means of Bayesian inference. The  likelihood, ${\mathcal L}( d | \bm{\theta})$, of the data $d$ given the parameters $\bm{\theta}$ of the source is~\cite{1994PhRvD..49.2658C}
\begin{equation}
\ln {\mathcal L}(d | \bm{\theta}) = -\!\sum_{k} \frac{\langle d_k - h_k(\bm{\theta}) | d_k - h_k(\bm{\theta}) \rangle_k}{2}\!+\mathrm{const},
\label{eq:like}
\end{equation}
where $h_k$ is the TDI output $k$ produced by the GW $h(t; \bm{\theta})$, or, equivalently, in the Fourier domain, $\tilde h(f; \bm{\theta})$.
The inner-product is defined as
\begin{equation}
    \langle a | b \rangle_k = 2 
    \int_0^{+\infty}\!
    \mathrm{d}f\; \frac{\tilde a(f) \tilde b^*(f) + \tilde a^*(f) \tilde b(f)}{S_k(f)} \,, \label{eq:innerprod}
\end{equation}
where $\tilde a(f)$ is the Fourier transform of the time series $a(t)$,  $S_k(f)$ is the noise power spectral density of the $k$th data stream, and the extrema $[f_\mathrm{low}, f_\mathrm{high}]$ corresponds to the frequency range spanned by a GW with parameters $\bm{\theta}$ over the duration of the observation.

Once a prior $p(\bm{\theta})$ is specified, we compute the joint posterior probability density function (PDF) on the parameters of the source
\begin{equation}
    p(\bm{\theta} | d) \propto {\mathcal L}\left(d | \bm{\theta}\right) p(\bm{\theta})
    \label{eq:posterior}
\end{equation}
through stochastic sampling. \textsc{Balrog} is designed to work with different sampler flavors and implementations. For the analysis presented here we use a nested sampling algorithm based on \textsc{CPNest}  \cite{john_veitch_2021_4470001}.

We model the gravitational waveforms $h(t; \bm{\theta})$ in their adiabatic inspiral regime through a post-Newtonian (PN) expansion, using two different waveform models. One of them is a new implementation under active development~\cite{2021arXiv210610291K} which
includes both spin precession and orbital eccentricity. Improving upon previous work~\cite{2018PhRvD..98j4043K},
the new formulation is substantially more efficient in terms of computational requirements, making the analysis presented here possible.
We also use a 3.5PN \textsc{TaylorF2} waveform (see e.g.~\cite{2001PhRvD..63d4023D,2005PhRvD..72b9901D}) restricted to aligned spins and quasicircular orbits when analyzing the \ldc{} dataset, in agreement with the signals injected in it. The full set of  waveforms we use are computed at sufficiently high PN order to ensure that no systematic effect is introduced in the analysis~\cite{2019PhRvD..99f4056M}.
We describe the TDI outputs from such a model as in Ref.~\cite{2018arXiv180610734M}, which allows us to fully reproduce the waveforms used in \ldc{}.
Each source is described by 17 (11) parameters in the precessing and eccentric (spin-aligned and circular) case.

\subsection{Implementation}
\label{subsec:implementation}

The noise-orthogonal TDI outputs $d$ on which the LISA GW analysis is based need to be computed from intermediate TDI data products, e.g. the TDI Michelson observables $X$, $Y$, and  $Z$~\cite{2004PhRvD..69b2001S,2005PhRvD..71b2001V}.
Note that the noise-orthogonal data channels $A$, $E$, and $T$ first introduced in the literature~\cite{2002PhRvD..66l2002P} were constructed from the Sagnac variables $\alpha$, $\beta$, and $\gamma$ and are therefore slightly different from the ones we are using here.
Here for concreteness and to consistently interface with the data currently generated within the LDCs, we start from $X$, $Y$, and  $Z$. This step will need to be revised in the future as the interplay between the raw phase-meter data and the actual GW analysis becomes clearer.

In order to improve our computational efficiency, we use a rigid adiabatic approximation (RAA) of the TDI variables~\cite{2004PhRvD..69h2003R}, that is approximately related to the 1.5-generation variables injected into the datasets as
\begin{align}
    \tilde{X}\sub{1.5-g}(f) &\approx \left( 1 - e^{-4 \pi i f L} \right) \tilde{X}\sub{RAA}(f),
\end{align}
where $L = 2.5 \times 10^9$~m is the mean LISA armlength,
and similarly for the other two TDI variables $Y$ and $Z$.
We note that \source{} sources accumulate most of their SNR at the high frequency end of the LISA bandwidth ($f\gtrsim 5\,\mathrm{mHz}$, where $fL\gtrsim 1$; see Fig.~\ref{fig:Catalog}). Therefore, a long-wavelength approximation to the detector response is not appropriate.
We note that the RAA is not faithful to the full TDI response at very high frequencies. Since we recover source parameters from full TDI signals with RAA signals, this study also serves as a test of the RAA for \source{} signals.

In order to compute inner products [cf. Eq.~\eqref{eq:innerprod}] involving a discrete time series, we use a hybrid method based on Clenshaw-Curtis quadrature.
First, the time series representing the data (having a cadence of 5~s in the \ldc{} case) is related through a discrete Fourier transform to a frequency series from $f\sub{min} = 0$ to $f\sub{max} = 0.1$~Hz, with a resolution of $\Delta f = 1/T\sub{obs} \approx 1.27 \times 10^{-8}$~Hz.
This defines a finite set of data points in the Fourier domain $f_i\super{DFT}$ with $f\sub{min} \leq f_i\super{DFT} \leq f\sub{max}$.
We transform the frequency interval into a log-frequency interval, and
split the latter into ten subintervals of equal length.
In each of them, we compute a 21-point Clenshaw-Curtis quadrature rule, resulting overall in a set of $N=201$ distinct frequencies $f_i\super{CC}$ with corresponding weights $w_i\super{CC}$.
For each of these points, we then find the closest frequency in the discrete set $f_i\super{DFT}$ to form the set $f_i\super{H}$. In order to construct the associated weights $w_i\super{H}$, we first note that each frequency $f_i\super{CC}$ satisfies either $f_i\super{CC} < f_0\super{H}$; $f_i\super{CC} > f_{N-1}\super{H}$; or $f_k\super{H} \leq f_i\super{CC} \leq f_{k+1}\super{H}$. In the first case, we associate the weight $w_i\super{CC}$ with $w_0\super{H}$; in the second case we associate the weight $w_i\super{CC}$ with $w_{N-1}\super{H}$; and in the third case we distribute the weight $w_i\super{CC}$ linearly (in log) between $w_k\super{H}$ and $w_{k+1}\super{H}$ according to the respective distance to $f_i\super{CC}$ of their corresponding frequencies.
Finally, some frequencies in the set $f_i\super{H}$ might be duplicates, in which case we combine them and their weights for minor gains in computational efficiency.
This results in a set of $N_H \leq N$ hybrid frequencies and weights allowing us to approximate the integral as
\begin{align}
    \int_a^b \mathrm{d}f\; g(f) \approx \sum_{k=0}^{N_H-1} w_k\super{H} g(f_k\super{H}).
\end{align}
We verified that the loss of accuracy in the integral evaluation due to the modification of the quadrature rule does not impact the result significantly.
With this method, we drastically reduce the number of waveform evaluations necessary to evaluate each log-likelihood by selecting a few relevant datapoints in the discrete Fourier transform of the data.
Note that this algorithm needs only to be used once at the beginning of the run, and that the computational efficiency of the resulting run is independent of the length of the time series, and weakly dependent on its cadence. We also stress that the choice of 10 subintervals with a 21-point Clenshaw-Curtis quadrature rule applied to them is somewhat arbitrary. We found that it yielded fast parameter estimation with good reliability in the case at hand, but it can certainly be optimized depending on the particular source analyzed.

\subsection{Sampling parameters}

\label{subsec:samplingparams}

An appropriate choice of the sampling parameters is crucial to complete the inference. In order to remove the influence of uncertain cosmological effects from the analysis, in what follows we express all mass parameters in the detector frame (i.e. we use redshifted masses), unless stated otherwise. We choose to use the following set of 11 parameters to describe the circular, spin-aligned \source{}s:
\begin{itemize}
    \item For the two mass parameters, we use the chirp mass $\mathcal{M}_c$ and the dimensionless mass difference $\delta\mu = (m_1 - m_2)/(m_1 + m_2)$.
    \item The two amplitude parameters $A_L$ and $A_R$ are related to the luminosity distance $D_L$ and the inclination $\iota$ by $A_L = (1 + \cos \iota)/\sqrt{2 D_L}$ and $A_R = (1 - \cos \iota)/\sqrt{2 D_L}$. They are the square roots of the amplitudes of the left- (right-)handed components of a GW.
    \item The two phase parameters $\psi_L$ and $\psi_R$ are related to the initial orbital phase $\phi_0$ and the polarization phase $\psi$ by $\psi_L = \phi_0 + \psi$ and $\psi_R = \phi_0 - \psi$. These are the initial phases of the left- (right-)handed components of a GW.
    \item The spins are described by the parameters $\chi_{1,\ell}$ and $\chi_{2,\ell}$, corresponding to the dimensionless spin magnitudes of the two binary components.
    \item The initial orbital frequency of the source $f_0\super{orb}$, related to the initial GW frequency by $f_0 = 2 f_0\super{orb}$.
    \item The sine of the ecliptic latitude $\sin b$ and the ecliptic longitude $l$, as sky location parameters.
\end{itemize}
For the case of eccentric sources with precessing spins, one needs to modify and extend the sampled set of parameters. In particular, we choose the following:
\begin{itemize}
    \item We parametrize eccentric orbits with the square eccentricity at $f=10$~mHz $e^2_{10}$ and the initial argument of periastron $\phi_e$.
    \item Precessing sources require six spin parameters; we choose the dimensionless spin magnitudes $\chi_{1,2}$ and the spin orientations in an ecliptic frame, which we describe by their sine latitudes $\sin b^\chi_{ 1,2}$ and longitudes  $l^\chi_{1,2}$.
    \item For these runs, we also use the approximate time to merger $t_M$ defined in Eq.~\eqref{eq:eccentricmergertime} instead of the initial orbital frequency $f_0\super{orb}$. 
\end{itemize}

We use flat priors for all the above parameters.
Assuming that some information on the source will be provided by the preceding search stage, we restrict the prior range for (at least some of) the parameters around the injected values, which is essential to keep the computational burden at a manageable level (cf. Secs.~\ref{subsec:challenge_results} and~\ref{subsec:eccprec}).

\begin{figure}[t]
    \centering
    \includegraphics[width=\columnwidth]{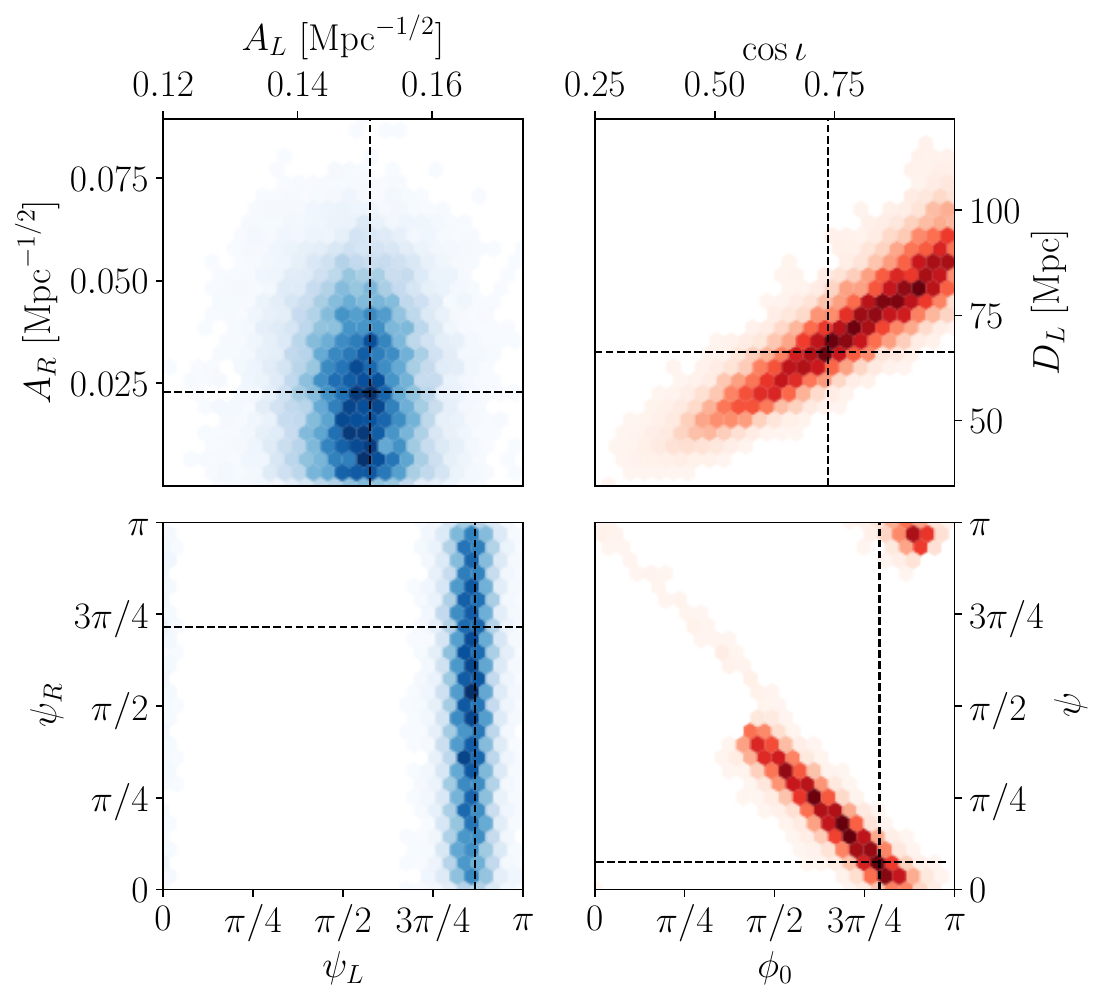}
    \caption{Two-dimensional posteriors for \ldc{} source 15 (SNR 12) using different parameters. Quantities related to the circular (linear) polarizations are indicated in blue (red). The amplitudes $A_L$ and $A_R$ (top left) are related to $D_L$ and $\cos \iota$ (top right). The phases  $\psi_L$ and $\psi_R$ (bottom left) are related to $\phi_0$ and $\psi$ (bottom right). The posteriors have been weighted in each plot so that the parameters shown in it are flat distributed.
    Quantities on the left (blue) are significantly less correlated than those on the right (red).}
    \label{fig:2x2corner}
\end{figure}

This specific choice of parameters greatly simplifies the likelihood  structure, thus facilitating the sampling process.
We use the chirp mass $\mathcal{M}_c$ because this is the mass parameter entering  the frequency evolution at lowest PN order and is thus better constrained than any other mass parameter.
For the second mass parameter, our choice of $\delta \mu$ has a key advantage over the more traditional alternatives of the symmetric mass ratio $\nu$ or the mass ratio $q$: the Jacobian of the transformation into the $(m_1,m_2)$ space is symmetric and regular in the $m_1 = m_2$ limit, avoiding potential issues related to this reparametrization.
As shown in Fig.~\ref{fig:2x2corner}, the amplitude parameters $A_L$ and $A_R$ are weakly correlated, in contrast with the more common choices of luminosity distance $D_L$ and inclination $\cos \iota$. Furthermore, a flat distribution in $A_L$ and $A_R$ corresponds to a flat distribution in $\cos \iota$, which we expect from an isotropic distribution of source locations and orbital angular momenta.
Similarly, in the interest of avoiding strongly correlated quantities, we opted for the phase parameters $\psi_L$ and $\psi_R$ instead of $\phi_0$ and $\psi$.

Figure~\ref{fig:2x2corner} shows a comparison of the two-dimensional posteriors for an illustrative \ldc{} run (source 15), which has signal-to-noise ratio (SNR) 12. We contrast the parameter spaces described by $(A_L,A_R)$ and $(D_L, \cos\iota)$ as well as that described by  $(\psi_L,\psi_R)$ and $(\phi_0,\psi)$.
The posterior distributions of parameters related to the circular polarization of the GW are
significantly less correlated compared to those involving linear polarization.
Moreover, 
we only sample half of the $(\phi_0,\psi)$ plane, thus removing the multimodality arising from the following symmetries of gravitational radiation:
$(\phi_0 \to \phi_0 + n\pi)$, $(\psi \to \psi + n \pi)$, $(\phi_0 \to \phi_0 + \pi/2, \psi \to \psi + \pi/2)$, $n \in \mathbb{Z}$.
Note that the source we chose for illustrative purposes offers an unbiased measurement of the phases $\psi_{L,R}$, while we observed significant biases in the recovery of those two parameters for most sources.
However, we did not observe such biases when analyzing data containing a single source, and we thus argue that this effect arises from the confusion between overlapping sources and is independent of the chosen parametrization. This illustrative source is thus representative of the single source injection results, and most relevant to this discussion.

%%%
\section{Results}
\label{sec:results}

As an initial test of the analysis approach described in the previous section, we have applied it to the data sets released for \ldc{}. The data sets are briefly described in Sec.~\ref{subsec:challenge}. Details of prior choices and parameter estimation results are presented in Sec.~\ref{subsec:challenge_results}. As \ldc{} had limited scope and contained only BHs on circular orbits with aligned spins, we also present in Sec.~\ref{subsec:eccprec} a proof-of-concept analysis on a generic precessing and eccentric system.

\subsection{LISA data challenge}
\label{subsec:challenge}

The first round of the LISA Data Challenges (\ldc{}, \href{https://lisa-ldc.lal.in2p3.fr/}{lisa-ldc.lal.in2p3.fr}) consisted of several datasets, each of which was dedicated to a specific source class: massive black hole binaries, extreme-mass ratio inspirals, galactic binaries, SmBBHs, and stochastic backgrounds.

The \ldc{} SmBBH data consists of two sets, each containing the same 66 \source{} injections, one noise-free and the other including a realization of the expected LISA Gaussian stationary noise.
In this work, we focused on the noiseless dataset, as our initial goal is to test the performance of the Bayesian analysis scheme to accurately recover the source parameters.
We are currently developing functionalities to jointly estimate the unknown level of noise that affect the source measurements, which we will report about in the future.
Each \ldc{} dataset consists of the 1.5-generation TDI observables $X$, $Y$, and  $Z$~\cite{2004PhRvD..69b2001S,2005PhRvD..71b2001V} with a cadence of 5 seconds and a duration of 2.5 years. By linearly combining $X$, $Y$, and $Z$, we construct the data, $d$, for the GW analysis, consisting of the three noise-orthogonal TDI observables $A$, $E$ and $T$~\cite{2002PhRvD..66l2002P}.

\begin{figure}[tbp]
    \centering
     \includegraphics[width=\columnwidth]{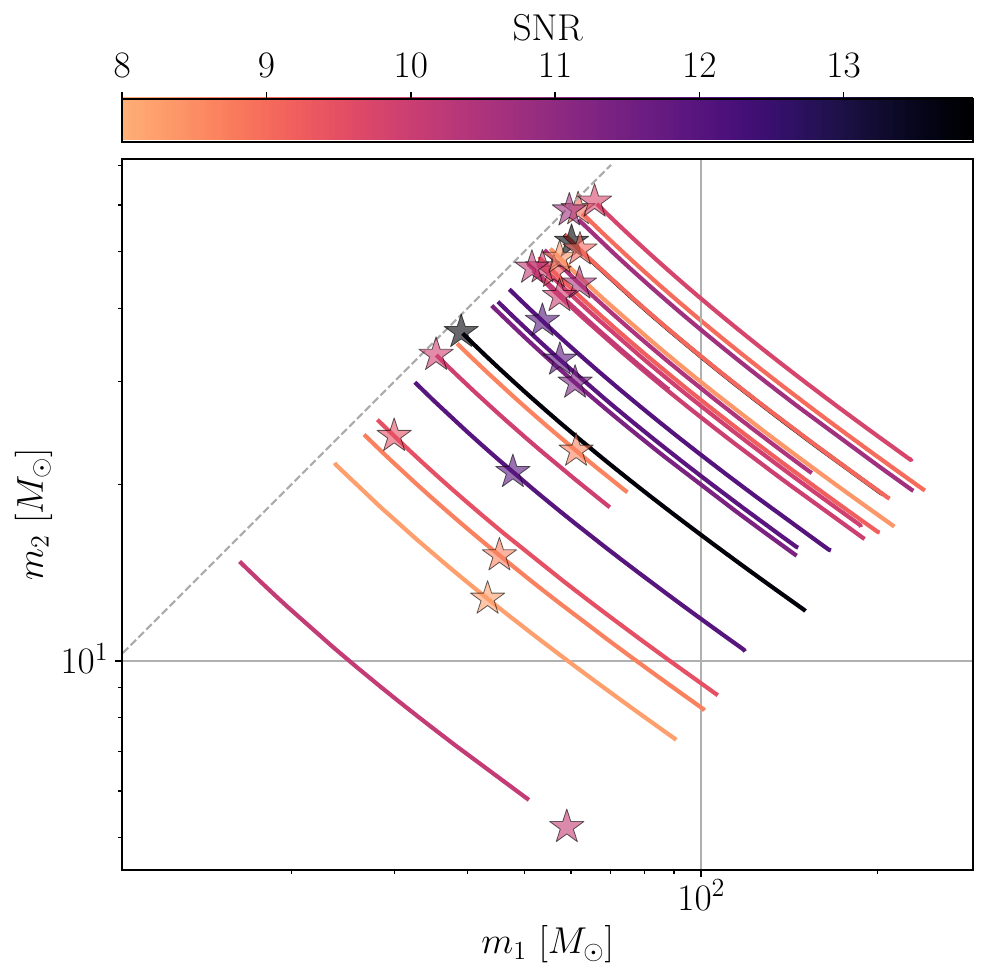}
    \caption{
    Posterior samples of detector frame component masses for the 22 recovered sources.
    Solid lines extend in the $90\%$ confidence interval of the symmetric mass ratio posterior.
    The thickness of the curve is comparable to or greater than the posterior distribution widths, indicating the very high accuracy of the chirp-mass measurements. Injected values are marked by stars.
    Lines and markers are colored according to the sources' SNRs.
    All injected values lie within their posterior's $90\%$ contour levels, except for source 16 ($\textrm{SNR}=10$) whose true, high dimensionless mass difference is within the $98\%$ confidence interval (cf. Sec.~\ref{subsec:triage}).}
    \label{fig:CompMasses}
\end{figure}

\begin{figure*}[tbp]
    \centering
    \includegraphics[width=1.0\textwidth]{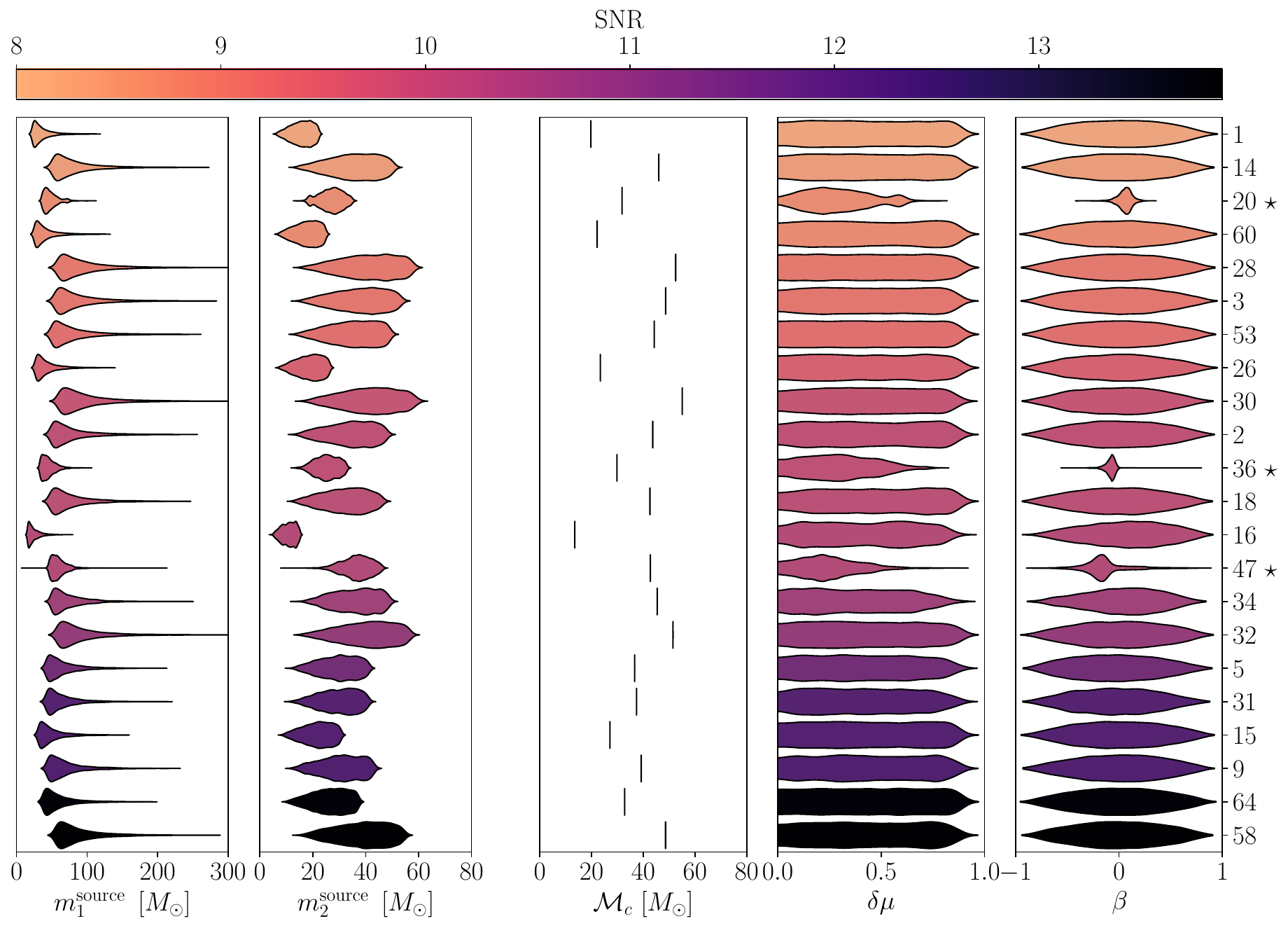}
    \caption{Marginal posteriors, represented through kernel density estimators for five selected parameters. From left to right, we show the primary component mass, the secondary component mass (both measured in the source frame),    
    the redshifted chirp mass, the dimensionless mass difference, and the 1.5 PN spin parameter reported in Eq.~\eqref{eq:beta}. Source posteriors are sorted and colored by their SNRs; their index in the LDC-1 injections catalog is reported to the right. Stars mark sources that merge within the \ldc{} dataset duration (2.5 yr).
    Posteriors are reweighted to an effective prior uniform in the column's parameter, except for the spin parameter $\beta$.
    The redshifted chirp mass, appearing in the leading-order PN term of the frequency evolution, is much better constrained than any of the other parameters. Parameters entering at higher PN order like $\delta\mu$ and $\beta$ can only be constrained for systems that merge within the mission lifetime.
    }
    \label{fig:ViolinCompMassesRich}
\end{figure*}

\begin{figure*}[tbp]
    \centering
    \includegraphics[origin=c, width=.9\textwidth]{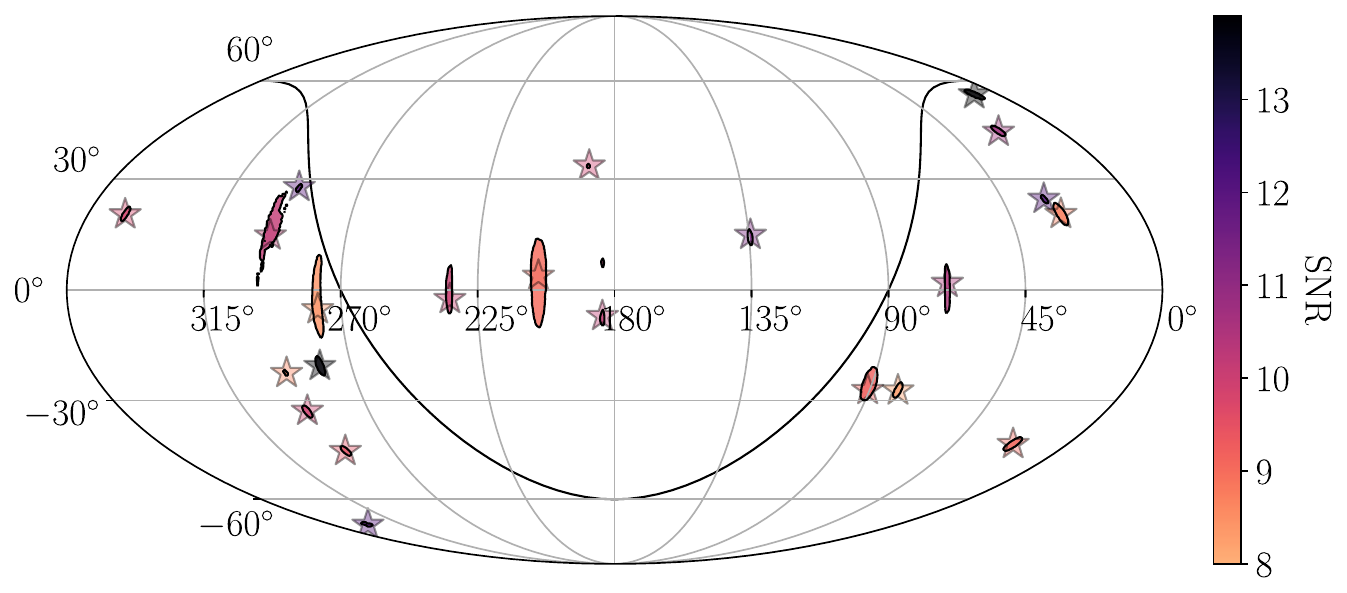}
    \caption{Posteriors on the sky position of the recovered sources in ecliptic coordinates using a Mollweide projection. Stars denote the true locations of the injected sources. All true locations are enclosed within the $90\%$ confidence intervals of their posterior. The solid black line shows the galactic plane.
    Note that sources close to the ecliptic have an approximate symmetry involving the ecliptic latitude $b \to -b$ (see e.g.~\cite{2021PhRvD.103h3011M}), resulting in elongated posteriors in that region, and even a bimodal posterior as seen in the source close to $180^\circ$ longitude.}
    \label{fig:Sky}
\end{figure*}

The parameters of the injected sources are released as part of the dataset. Figure~\ref{fig:Catalog} provides a summary of the main features of the signals that were injected.
These sources all have GW frequencies of $\sim 1$--$10\,\mathrm{mHz}$ at the beginning of the LISA mission.
This set of sources covers the chirp mass range $7 - 61$ $M_\odot$, see 
Figs.~\ref{fig:CompMasses},~\ref{fig:ViolinCompMassesRich}.
The source chirp masses and initial frequencies determine the merger time [see Eq.~\eqref{eq:mergertime}].
Five sources
inspiral and merge within $T_\mathrm{obs} = 2.5~\mathrm{yr}$,
with chirp masses in the range $30$--$61 M_\odot$
and initial frequencies between 16 and 20 mHz.
Five more,
with chirp masses in the range $20$--$47 M_\odot$ and frequencies between 12 and 22 mHz
merge within 5 years.
Six other,
with chirp masses in the range $13$--$55 M_\odot$ and frequencies between 8 and 21 mHz
merge within 10 years.
The longest lived ones,
with chirp masses in the range $7$--$55 M_\odot$ and frequencies between 1 and 11 mHz
merge within $3000$ years.
This set of sources covers a range of SNRs which is governed primarily by the distance to the source, the inclination angle and the source sky position, the latter of which is shown in Fig.~\ref{fig:Sky}. In total, 22 sources yield an optimal (and coherent across the 3 TDI observables) $\text{SNR}>8$.

\begin{figure}[t]
    \centering
    \includegraphics[width=\columnwidth]{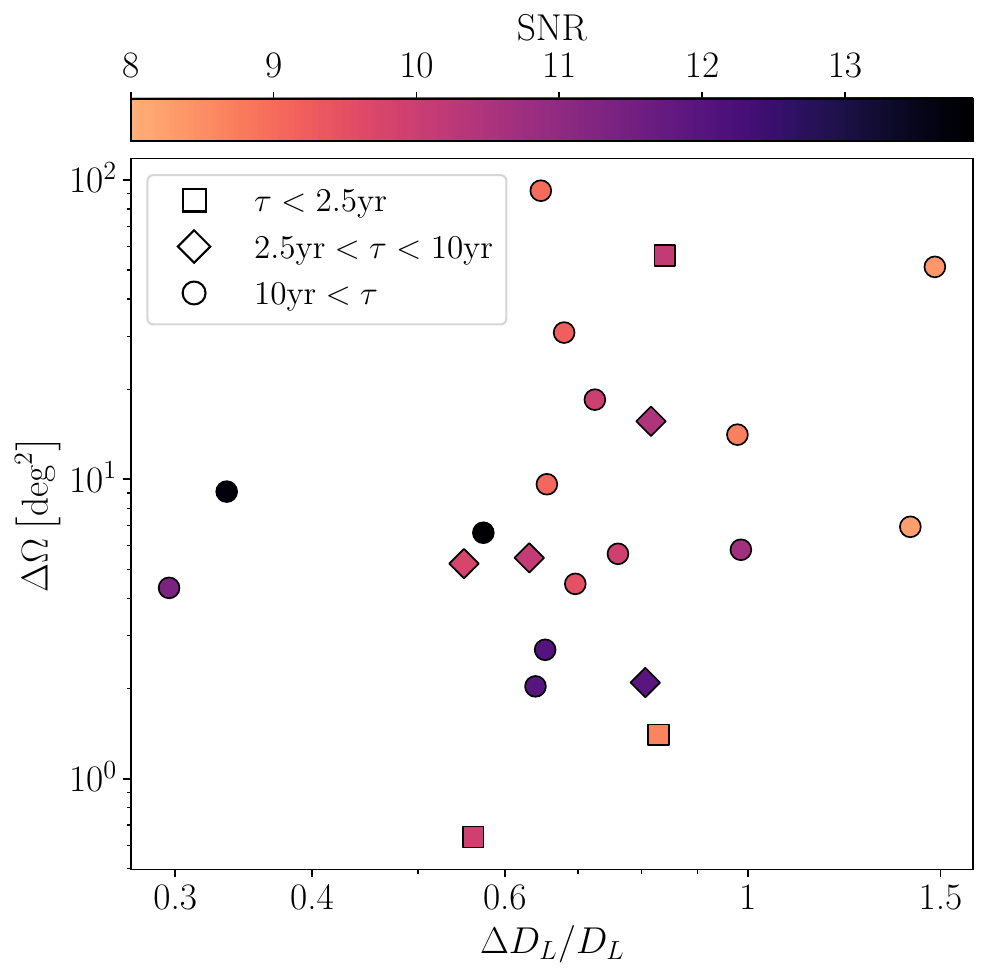}
    \caption{
    Marginalized posterior uncertainties on the distance and sky location of the 22 recovered sources. We show  the $90\%$ confidence intervals of the two--dimensional sky location posteriors and the uncertainties on the distance relative to its true injected value.
    All sources above $\text{SNR}>8$ are resolved with angular resolution better than $100~\text{deg}^2$.
    At $\text{SNR}$ above $11$ the localization improves by an order of magnitude and the distance is measured with $70\%$ accuracy or better.
    Markers label the source time to coalescence as shown in legend and described previously in Fig.~\ref{fig:Catalog}.}
    \label{fig:Localization}
\end{figure}

\subsection{Parameter estimation and results}
\label{subsec:challenge_results}

Preliminarily, we analyzed the same noiseless dataset in distinct runs, where we tuned the priors to a corresponding target \source{}.
We chose priors
\begin{itemize}
    \item Flat in $\delta\mu$ in  $[0,0.9]$, corresponding to a mass ratio between $1:1$ and $1:19$.
    \item Flat in $A_L$ and $A_R$ in $[0, A\sub{max}]$, where $A\sub{max}=2\sqrt{2/D_L}$ (twice the overall amplitude of an optimally oriented source at the injected distance).
    \item Flat in $\psi_L$ and $\psi_R$ in $[0, \pi]$.
    \item Flat in $\chi_{1,\ell}$ and $\chi_{2,\ell}$ in $[-1, 1]$.
    \item For $\mathcal{M}_c$, $f_0$, $\sin b$, and $l$, we emulated the output of a prior GW search by performing searches on single source simulated data in steps.
    At each step, we adjust the priors using the posteriors resulting from the previous step to $m \pm 4\sigma$, where $m$ is the median of the posterior, and $\sigma$ its standard deviation.
    In order to improve the convergence of the method, when computing $m$ and $\sigma$ we neglected posterior samples with log-likelihood smaller than one obtained with $A_L=A_R=0$ (i.e. with $\log \mathcal{L} < -\text{SNR}^2/2$).
    This method required at most three steps for each target before convergence.
    Note that this was not possible for all systems, particularly those with low SNR, which we flagged as not detected.
\end{itemize}
The method we used to determine the search priors produced a set of single-injection runs, where the same waveforms were used for injection and recovery. We found it a useful set of analyses to compare to our main results.

\begin{figure}[tbp]
\centering
\includegraphics[width=\columnwidth]{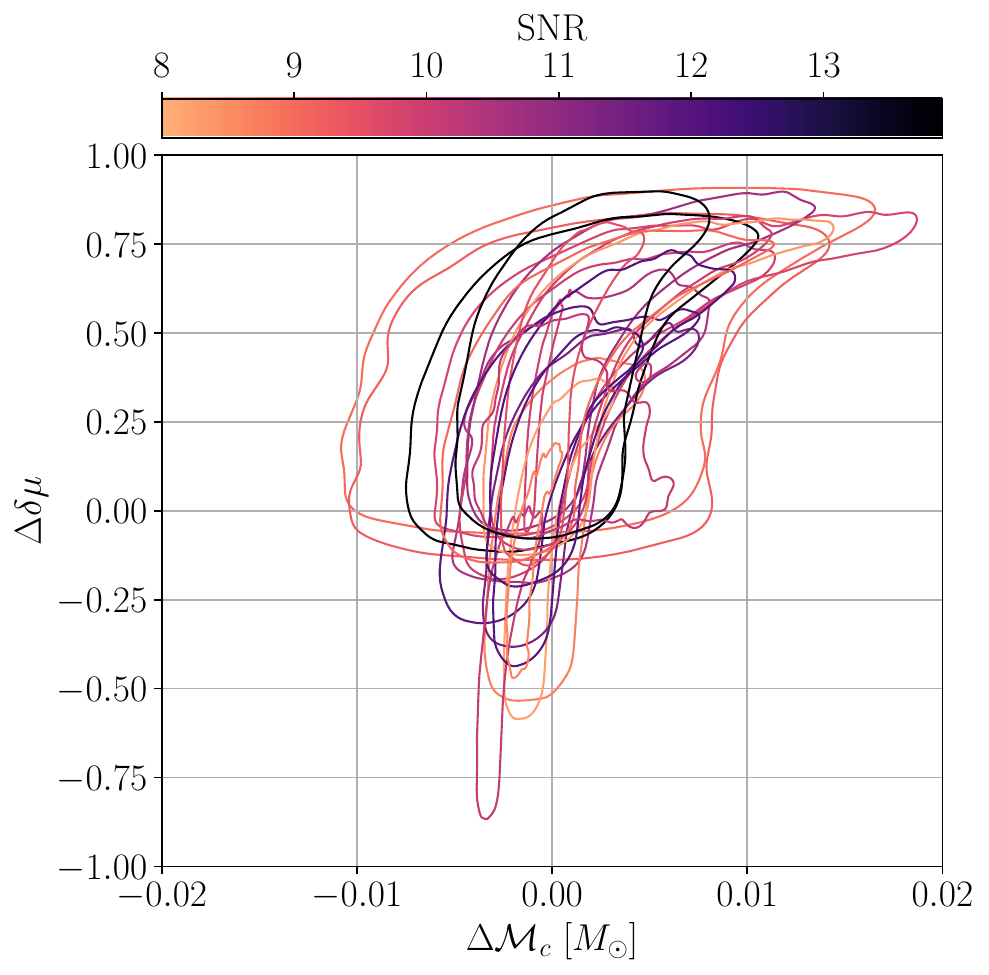}
\caption{
90\% marginalized posterior contours in the chirp mass-dimensionless mass difference plane for all 22 observed sources.
The colors of the contours are set by the SNR of the corresponding source. All contours are offset so that the injected value of the corresponding source lies at the origin.}
    \label{fig:Mchirp}
\end{figure}

This set of sources also covers a range of SNRs which is governed primarily by the distance to the source, the viewing inclination angle and the source sky position in addition to the chirp mass.
With our parameter-estimation pipeline, we were able to obtain good quality posterior distributions, and hence measurements of the source parameters for the 22 sources with $\text{SNR}>8$.
Eight sources with $\text{SNR}$s in the range 5.7--7.9 offered good quality posteriors as well, but we choose to use a fixed SNR threshold and exclude them from the analysis.

For each of the $22$ sources that we selected, we computed the 3-dimensional volume within which LISA is able to localize it.
Because these sources are generally long-lived, and are at the high-frequency end of the LISA bandwidth, relatively good (by the standards of GW astronomy) sky position measurements with uncertainty regions spanning $\Delta\Omega=1$ to $100$ square degrees are obtained.
However, because these sources have relatively low SNRs in the range 8--14 there is a comparatively large fractional uncertainty in the distances spanning 30\%--150\%.
These results are summarized in Figs.~\ref{fig:Sky} and \ref{fig:Localization}.

Of the intrinsic source parameters, by far the best measured is the chirp mass;
for the loudest (quietest) of the recovered sources with SNR 14 (8)
we find that we are able to measure the chirp mass to a fractional accuracy better than 0.5\% (2\%).
Our parameter-estimation pipeline sampled directly in the chirp mass $\mathcal{M}_c$ and the dimensionless mass difference $\delta\mu$ as explained in Sec.~\ref{subsec:samplingparams}.
The resulting posteriors are shown in Fig.~\ref{fig:Mchirp}.
The more astrophysically interesting component masses $m_1$ and $m_2$ for the individual BHs can be obtained from $\mathcal{M}_c$ and $\delta\mu$;
see Fig.~\ref{fig:CompMasses}.
Notably, we find fractional uncertainties on chirp masses ---measured in the frame at rest with the Hubble flow--- to be comparable or smaller ($\Delta \mathcal{M}_c^{\mathrm{H}}/\mathcal{M}_c^{\mathrm{H}}\lesssim 2\times 10^{-2}$) to the uncertainties arising from source proper motion redshifts ($v_{\mathrm{pec}}/c \lesssim 10^{-2}$).
Similarly, the choice of cosmology yields uncertainties in redshift up to $10^{-2}$ for the most distant source recovered at $500~\mathrm{Mpc}$, over a broad range of cosmological parameters~\cite{2020A&A...641A...6P,2021ApJ...908L...6R}.

Of the other intrinsic parameters, the most interesting are arguably the component spins. While $\chi_{1,\ell}$ and $\chi_{2,\ell}$ cannot be individually measured, it is helpful to identify intrinsic parameters entering the PN frequency evolution series at different orders~\cite{2014LRR....17....2B,1993PhRvD..47.4183K,1994PhRvD..49.2658C}. As mentioned above, the parameter entering the series at leading order is $\mathcal{M}_c$, the parameter entering at 1PN is $\delta\mu$, while spins first enter at 1.5PN via the combination 
\begin{align}
    \beta &= \sum_{i=1}^2 \left( \mu_i + \frac{75 \mu_j}{113} \right) \mu_i \chi_{i,\ell}, \label{eq:beta}
\end{align}
where $i \neq j$, and $\mu_i = m_i/(m_1 + m_2)$ are the dimensionless individual masses. We normalized this parameter so that $|\beta| \leq 1$ for arbitrary mass ratios, implying $|\beta| \leq 94/113$ for equal-mass systems.

The marginal posterior distributions of these three parameters are shown in Fig.~\ref{fig:ViolinCompMassesRich}, together with those of the individual masses $m_1$ and $m_2$. While the chirp mass is measured extremely well for all sources, $\delta\mu$ and $\beta$ can be measured with some confidence only for SmBBHs that are merging within the observation window. This is because those sources are the only ones with a sufficient frequency evolution such that the subdominant terms in the PN expansion become observable.

Overall, comparing runs performed on the \ldc{} data with ones performed on single-source injections, we find that parameters are recovered with similar precision. Biases in the \ldc{} runs are comparable to those expected from random noise fluctuations.
The exception to that are the two phase parameters $\psi_L$ and $\psi_R$. These were recovered without any significant bias in the single-source runs,
but with large biases comparable to the prior range in the \ldc{} runs in almost all cases. These biases did propagate to both parameters when converted to the $(\phi_0, \psi)$ plane.

We summarize in Table~\ref{tab:injection} of the appendix the injected parameters of the 22 sources we analyzed, and in Table~\ref{tab:recovery} their recovered values.

\subsection{Challenging systems}
\label{subsec:triage}

Let us now discuss those few systems which showed posteriors that were found to be particularly challenging to analyze.
All the \source{} injections and recoveries done in \ldc{} were performed using noiseless injections.
Therefore, in the absence of noise fluctuations, we might expect the likelihood (posterior) to be peaked at (near) the true (i.e. injected) source parameters.
However, this is not guaranteed to be the case because (i) we are using different waveforms for recovery than the ones that were injected,
and (ii) some sources are overlapping in the \ldc{} data and could therefore be confused.

In particular, we highlight four systems. 
\begin{itemize}
    \item For source number 5 ($\text{SNR}=11.36$), we obtain a frequency posterior that is peaked significantly away from the injected values.
    \item Sources number 20 ($\text{SNR}=8.68$), and 36 ($\text{SNR}=9.93$) resulted in a 2-dimensional posterior on the chirp mass and mass difference parameters (or equivalently on the component masses) that only include the injected values on the boundary of their $\sim 99.8\%$ confidence interval.
    \item For source number 16 ($\text{SNR}=10.14$) the marginalized, 1-dimensional posterior on $\mathcal{M}_c$ includes the injected value only in its 99.4\% confidence interval.
\end{itemize}

We note that for the first two bullet points listed above, the issues described are not present in the single-injection run results used for comparison.
These differences could possibly be due to the difference in the employed waveforms, the signal overlap in \ldc{}, sampling issues, or a combination of these. 
Work toward analyzing jointly the overlapping sources~\cite{2019PhRvD.100h4041B} and characterizing performances of different samplers is ongoing.

\begin{figure}[htbp]
    \centering
    \includegraphics[width=\columnwidth]{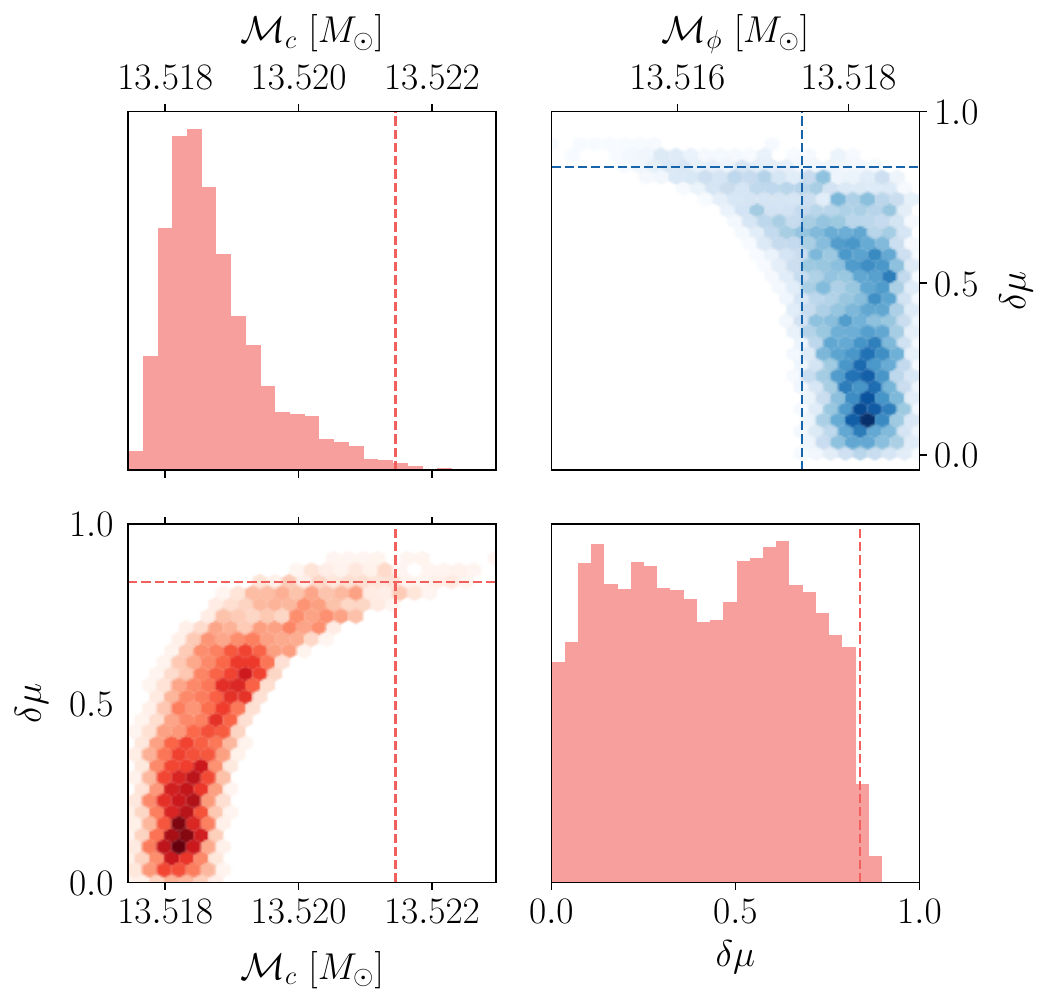}
    \caption{
    Marginalized posterior distributions in the masses plane for source number 16.
    The two-dimensional posterior for $(\mathcal{M}_c,\delta\mu)$ (red-shaded, lower left plot) shows support in the region containing the true injected value (dashed red line).
    However, the correlation structure in these parameters together with the fact that $\delta\mu$ is not measurable generate a strong bias in $\mathcal{M}_c$ when marginalized over $\delta\mu$, as shown in the upper left and lower right histograms. The corresponding plot in the single-source injection run showed a similar pattern.
    By re-parameterizing the masses plane with $(\mathcal{M}_\phi,\delta\mu)$ (blue shaded, upper right plot), we observe a milder correlation, hence a smaller bias on the posterior marginal distribution for $\mathcal{M}_\phi$ compared to $\mathcal{M}_c$.
    }
    \label{fig:massescorner16}
\end{figure}

On the other hand, the bias in $\mathcal{M}_c$ observed for source 16 was also present in the single-injection result. In the following we argue that it's a genuine effect of the signal parametrization.
Figure~\ref{fig:massescorner16} shows the marginalized posteriors in the $(\mathcal{M}_c,\delta\mu)$ plane for source number 16 ($\text{SNR}=10.14$, $\tau=7.9$~yr), together with a reparametrization of it.
While the true parameters still lie in the main confidence region of the two-dimensional posterior, the chirp mass posterior only includes the injected value in the tail of the distribution. Notably, its injected mass ratio of $q \approx 1/11.3$ ($\delta\mu \approx 0.84$) is the most asymmetric among all detected sources.  The flat posterior in $\delta\mu$ that we observe in Fig.~\ref{fig:massescorner16} suggests that this parameter is not measurable. The flatness of this posterior together with the shape of the confidence region implies a bias in the marginalized posterior for $\mathcal{M}_c$ for highly asymmetric mass ratios.
The shape of the two-dimensional posterior can be explained by an examination of the PN GW phase series~\cite{2014LRR....17....2B}:
\begin{align}
 \Phi &= \Phi_c - \frac{(\pi \mathcal{M}_c f)^{-5/3}}{16} \bigg[ 1 \\
 &+ \frac{(\pi \mathcal{M}_c f)^{2/3}}{2^{1/5}\left( 1 - \delta\mu^2\right)^{2/5}} \left( \frac{2435}{252} - \frac{55 \delta \mu^2}{24} \right) \bigg] + \mathcal{O} \left(f^{-2/3}\right).\nonumber
\end{align}
As $\delta\mu$ increases, the resulting change in the number of accumulated cycles can be compensated by an increase in $\mathcal{M}_c$. This behavior is all the more pronounced that the system is observed closer to merger, as the strength of the 1PN term gets more comparable to the 0PN one. 
To reduce the correlation in the $(\mathcal{M}_c, \delta\mu)$ plane, we can define a new parameter 
$\mathcal{M}_\phi(\mathcal{M}_c, \delta\mu, f_0, T\sub{obs})$,
such that the number of accumulated cycles during the observation is independent of $\delta\mu$ up to some given PN order. 
Note that the likelihood is not exclusively determined by the number of accumulated cycles of phase, hence one should not expect $\mathcal{M}_\phi$ and $\delta\mu$ to be completely uncorrelated.
At 1PN order, we get
\begin{align}
    \mathcal{M}_\phi &= \mathcal{M}_c \Bigg\{1 - 
    \frac{(5 \mathcal{M}_c)^{1/4} \left[ 974 (1 - A) - 231 \delta\mu^2 \right]}{168\times 2^{1/5} A} \label{eq:Mphi}\\
    &\times \frac{\left( \tau_0^{3/8} - \tau\sub{f}^{3/8}\right)^2}{5\tau_0 -8 \tau_0^{5/8} \tau\sub{f}^{3/8} + 3 \tau\sub{f}}
    \Bigg\}, \nonumber\\
    A &= \left(1 - \delta\mu^2\right)^{2/5}, \\
    \tau_0 &= \frac{5 (\pi \mathcal{M}_c f_0)^{-5/3}}{256 \pi f_0}, \\
    \tau\sub{f} &= \max \left[ \tau_0 - T\sub{obs}, \frac{5 (\pi \mathcal{M}_c f\sub{max})^{-5/3}}{256 \pi f\sub{max}} \right],
\end{align}
where $f_0 = 2 f_0\super{orb}$ is the initial GW frequency, and $f\sub{max}$ is the higher limit of the observation frequency band.
As shown in Fig.~\ref{fig:massescorner16}, the 2-dimensional posterior in the masses plane  yields a milder correlation, and hence a smaller bias in the marginalized, 1-dimensional posterior for $\mathcal{M}_\phi$.

\subsection{Eccentric precessing system}
\label{subsec:eccprec}

We also ran as a proof of concept a Bayesian parameter estimation run on a fully general eccentric precessing system. We chose a $95$-$55 M_\odot$ binary system, with spin magnitudes $\chi_1 = 0.7$ and $\chi_2 = 0.73$ respectively, initial spin misalignment angles $\theta_1 = 179^\circ$ and $\theta_2 = 135^\circ$ respectively, eccentricity at $10$~mHz of $e_{10} = 3.1 \times 10^{-3}$, and SNR $15$. These values were inspired by the most massive event detected by LIGO/Virgo to date, GW190521 \cite{2020PhRvL.125j1102A}.
This particular source accumulated ${\cal N} \approx 1.89 \times 10^6$ cycles of orbital phase, ${\cal N}\sub{spin} \approx 892$ cycles of spin precession, and ${\cal N}\sub{ecc} \approx 4060$ cycles of periastron precession. 

For this run, we used the same sampling parameters as for the \ldc{} runs with a few modifications. We used different spin parameters, we added eccentricity parameters, and we replaced the initial orbital frequency with the approximate merger time parameter~\cite{1963PhRv..131..435P}
\begin{align}
 t_M &= t_0 + \frac{5 \mathcal{M}_c  (\pi \mathcal{M}_c f_0)^{-8/3}}{32 \sqrt{1 - e^2_{10}} \left( 8 + 7 e^2_{10}\right)}, \label{eq:eccentricmergertime}
\end{align}
where $t_0$ is the time at the start of data gathering.
Note that, for simplicity, this relation is obtained from the leading PN order frequency evolution equation, assuming a constant eccentricity. It is more accurate for circular systems, and becomes gradually less so as the initial eccentricity increases.

\begin{figure*}[t]
    \centering
    \includegraphics[width=2\columnwidth]{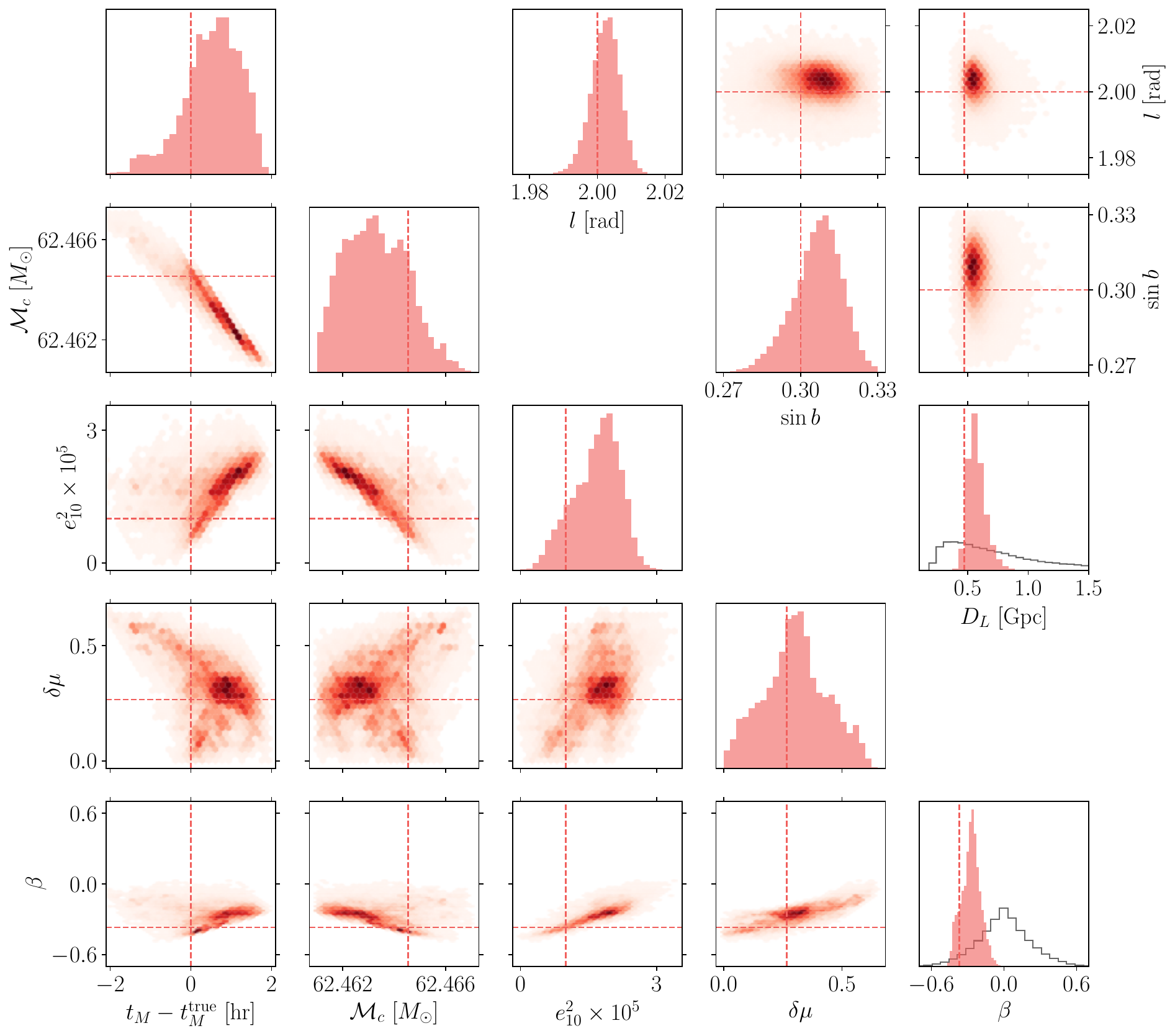}
    \caption{
    Posterior distributions on a selection of parameters of the eccentric precessing system. 
    Priors are uniform for (upper right panel) the ecliptic longitude $l$, the sine of ecliptic latitude $\sin b$, the amplitude parameters $A_L$ and $A_R$. Consequently, the luminosity distance prior is $\propto 1/D_L^2$, and shown as solid black line. 
    Equally, uniform priors are used for (lower left panel) the merger time $t_M$, the chirp mass $\mathcal{M}_c$, the square eccentricity $e_{10}^2$, the dimensionless mass difference $\delta\mu$, the dimensionless spin magnitudes, and the spin unit vectors on the sphere. The resulting prior is shown for the spin parameter $\beta$ as a solid black line. The merger time can be measured within approximately an hour, and the eccentricity and spin parameter can be well distinguished from zero. The source is localized in the sky within $8 ~\textrm{deg}^2$.}
    \label{fig:eccentric}
\end{figure*}

We note that since spin-induced precession causes $\cos \iota$ and $\psi$ to evolve with time, we use their initial values to define the parameters $A_L$, $A_R$, $\psi_L$, and $\psi_R$. Additionally, we set the priors on $\psi_L$ and $\psi_R$ as $\left[0, 2\pi\right]$ since the addition of eccentricity breaks the waveform symmetry from $(\phi_0 \to \phi_0 + n \pi)$ to $(\phi_0 \to \phi_0 + 2 n \pi)$, $n \in \mathbb{Z}$.

Of note, we report on the measurability of a few chosen parameters, shown in Fig.~\ref{fig:eccentric}. 
The merger time $t_M$ could be recovered with accuracy $\sim 2$~hours, a figure comparable to the corresponding merging circular sources. 
The chirp mass could be recovered with accuracy of $\sim 0.004 M_\odot$.
The eccentricity at 10~mHz could be recovered in the range $2.7 \times 10^{-3} < e_{10} < 4.9 \times 10^{-3}$ at 90\% confidence, and was clearly distinguishable from zero.
The injected dimensionless mass difference was recovered within the 24\% confidence interval.
The (initial) spin parameter $\beta$ could be recovered with 90\% confidence in the range $-0.41 < \beta < -0.14$, including the injected value of $\beta\sub{inj} \approx -0.37$, (with $\beta = 0$ excluded at more than 99.9\% confidence).
The source is located on the sky at 90\% confidence level within $8~\text{deg}^2$.
The recovered values were mostly consistent with the injected ones. More work is ongoing to assess the robustness of the parameter estimation pipeline across the full parameter space.

One interesting additional information to gather from these results is to determine whether the effects of spin-precession are measurable for such systems.
In order to do this, we looked at the average precession parameter $\chi_p$~\cite{2015PhRvD..91b4043S,2021PhRvD.103f4067G}, and found that the posterior did not differ significantly from the prior, suggesting that precession effects might not be measurable for \source{}s with LISA, thus strengthening the case for multiband GW astronomy.

These Bayesian results obtained for a fully precessing eccentric binary show promise for an extension of the present work, investigating the full 17-dimensional parameter space of \source{}s more extensively.

\subsection{Computational performances}
\label{cpuhsec}
Parameter estimation runs were carried out on the high performance computing infrastructure provided by the Birmingham BlueBEAR cluster, with each run using 8 Intel Xeon (2.50GHz) sibling cores on a single computing node.
The total CPU time for each run on the \ldc{} dataset was distributed with a median of 5 hours for the 22 sources with $\text{SNR}>8$.

The three runs with sources coalescing within the dataset duration where the most computationally demanding with CPU times of 36, 20, and 45 hours for sources number 20, 36, and 47 respectively.
All runs had small memory footprint throughout, with usage peaks below 1.6 Gigabytes.

The run with eccentricity and spin precession was substantially more expensive ($\sim 3000$~CPUh), approximately 100 times more than its merging spin-aligned circular counterparts. 
This is due to a combination of factors including the increased dimensionality of the parameter space,  the additional structure of the likelihood, and the increased complexity of the waveforms.

\section{Conclusions}
\label{sec:concl}
%%%

In this paper we presented a fully Bayesian parameter-estimation routine for the observation of  \source{}s with LISA.
As part of the LISA data challenge \ldc{}, we employed our codebase \textsc{Balrog} for the accurate estimation of 66 circular, spin-aligned \source{}s' parameters. We confidently recovered all 22 sources with $\text{SNR}>8$.
Our results show that LISA will be able to localize \source{}s over the sky within a few tens of squared degrees, and constrain their detector-frame chirp mass down to $\pm 0.01~M_\odot$.
Additionally, for sources merging within the mission lifetime, the chirping morphology of the signals allows us to measure parameters entering at higher order in the post-Netwonian expansion, namely the dimensionless mass difference $\delta\mu$, and the spin combination $\beta$.

On the technical side, we presented a novel choice of the sampling parameters that substantially reduce the correlations in the high-dimensional likelihood, thus vastly increasing the resulting computational efficiency. In particular, this relies on decomposing the signal into circular polarizations.
We also presented an algorithm that drastically reduces the number of waveform evaluations needed to estimate likelihoods, by adapting a nonuniform quadrature rule to work with uniformly sampled data. This allowed us to successfully perform full Bayesian parameter estimation studies for individual spin-aligned, circular \source{} sources undergoing $\mathcal{O}(10^6)$ wave cycles that required just a few CPU hours to complete.
Focusing on a selected number of sources that exhibit mild biases in the recovered parameters, we characterized the effect of binaries cross-contamination, waveform differences, and inherent likelihood structures which make \source{}s parameters challenging to sample.

Finally, we presented a proof-of-concept analysis where we tackle the full \source{} parameter-estimation problem, which includes eccentricity and spin precession. We recover the parameters of a specific, but generic, source in the resulting 17-dimensional parameter space.
We report a measurable eccentricity at 10~mHz of a few $10^{-3}$ together with a merger time determination within a time window of $\lesssim 1\,$ hour. 
We also report the immeasurability of spin-induced precession effects, suggesting that individual component spins cannot be recovered. This suggests that joint space and ground based detector GW observations might be crucial to fully characterize \source{}s.
More work is necessary to fully explore potential challenges for this type of sources.
This analysis brings us closer to LISA's goal of efficiently and accurately reconstructing the parameters of \source{}s, which constitute an unmatched tool to discriminate their formation history and evolution.

\acknowledgments
We thank
Christopher~P.~L.~Berry,
Sebastian~M.~Gaebel,
Janna~Goldstein,
Siyuan Chen,
Patricia Schmidt, and 
Geraint Pratten
for discussions.
We thank the members of the LISA Data Challenge working group for providing support and documentation on the dataset.
C.J.M. and A.V. acknowledge the support of the UK Space Agency through Grant No.  ST/V002813/1.
A.V. acknowledges the support of the Royal Society and Wolfson Foundation. D.G. is supported by European Union's H2020 ERC Starting Grant No. 945155--GWmining, Leverhulme Trust Grant No. RPG-2019-350, and Royal Society Grant No. RGS-R2-202004.  Computational work was performed on the University of Birmingham BlueBEAR cluster.
\textit{Software}:
We acknowledge usage of the following
\textsc{Python}~\cite{10.5555/1593511}
packages for the analysis, post-processing and production of results throughout:
\textsc{CPNest}~\cite{john_veitch_2021_4470001},
\textsc{ligo.skymap}~\cite{2016PhRvD..93b4013S},
\textsc{matplotlib}~\cite{2007CSE.....9...90H},
\textsc{seaborn}~\cite{2021JOSS....6.3021W},
\textsc{pandas}~\cite{reback2020pandas},
\textsc{pycbc}~\cite{alex_nitz_2021_4849433},
\textsc{ldc}~\cite{ldc_software},
\textsc{astropy}~\cite{2018AJ....156..123A},
\textsc{numpy}~\cite{2020Natur.585..357H},
\textsc{scipy}~\cite{2020NatMe..17..261V}.

\vfill

%%%
\bibliographystyle{apsrev4-2}
\bibliography{main.bib}
% Use this at submission on APS
%\input{output.bbl}
%%%
\clearpage

%%%%%%%%%%%%%%%%%%
%%% APPENDICES %%%
%%%%%%%%%%%%%%%%%%
\appendix
\widetext
%%%
\section*{Appendix: \ldc{} injected and recovered parameters}

In this Appendix, we provide the parameters of our \ldc{} analysis in tabular format. In particular, in Table~\ref{tab:injection} we list the parameters of all the injected \source{}s, while in  Table~\ref{tab:recovery}  we present the results of our parameter-estimation recovery. 

\begin{table}[ht]
\label{tab:injection}
\caption{Properties of the \ldc{} injected sources. Rows are ordered by increasing source SNR and labelled by the injection ID in the \ldc{} dataset. Sources merging within the mission lifetime (here set to 2.5 yr) are marked with stars. For a description of the parameters see Sec.~\ref{subsec:samplingparams}.}
\vspace{0.1cm}
\def\arraystretch{1.5}
\centering
\begin{tabular}{cccccccccccc|r}
\toprule
 $\mathrm{SNR}$ &  $\tau\,\left[\mathrm{yr}\right]$ &  $f_0\,\left[\mathrm{mHz}\right]$ &  $l\,\left[\mathrm{rad}\right]$ &  $\sin b$ &  $D_L\,\left[\mathrm{Mpc}\right]$ &  $\cos \iota$ &  $m_1\,\left[M_{\odot}\right]$ &  $m_2\,\left[M_{\odot}\right]$ &  $\mathcal{M}_c\,\left[M_{\odot}\right]$ &  $\delta\mu$ &  $\beta$ & $\mathrm{ID}$ \\
\midrule \midrule
           8.26 &                              45.7 &                            8.6020 &                            1.40 &     -0.46 &                              55.1 &          0.39 &                           43.2 &                           12.8 &                                   19.756 &         0.54 &     0.30 &             1 \\
           8.40 &                              27.5 &                            6.1377 &                            4.85 &     -0.09 &                             147.8 &         -0.28 &                           57.4 &                           48.7 &                                   46.016 &         0.08 &    -0.18 &            14 \\
           8.69 &                               2.3 &                           19.7430 &                            5.12 &     -0.38 &                             193.7 &          0.52 &                           61.2 &                           22.8 &                                   31.808 &         0.46 &     0.00 &     20$\star$ \\
           8.70 &                              84.3 &                            6.3652 &                            0.48 &      0.35 &                              68.0 &          0.62 &                           45.3 &                           15.1 &                                   22.158 &         0.50 &    -0.07 &            60 \\
           9.04 &                              89.5 &                            3.6289 &                            3.58 &      0.07 &                             237.2 &          0.99 &                           61.7 &                           59.0 &                                   52.539 &         0.02 &     0.13 &            28 \\
           9.07 &                              40.2 &                            5.1370 &                            0.38 &     -0.68 &                             238.7 &          0.84 &                           62.2 &                           50.4 &                                   48.659 &         0.10 &     0.18 &             3 \\
           9.21 &                             108.5 &                            3.7577 &                            1.58 &     -0.45 &                             180.8 &         -0.98 &                           56.1 &                           46.1 &                                   44.244 &         0.10 &     0.21 &            53 \\
           9.48 &                              51.7 &                            7.3862 &                            5.05 &     -0.70 &                              87.8 &          0.88 &                           30.0 &                           24.2 &                                   23.418 &         0.11 &     0.13 &            26 \\
           9.81 &                               8.5 &                            8.5140 &                            6.06 &      0.35 &                             493.4 &         -0.96 &                           65.8 &                           60.8 &                                   55.072 &         0.04 &     0.00 &            30 \\
           9.92 &                              23.9 &                            6.6834 &                            5.10 &     -0.54 &                             183.1 &          0.69 &                           53.6 &                           47.0 &                                   43.669 &         0.07 &    -0.14 &             2 \\
           9.93 &                               2.3 &                           20.4730 &                            3.30 &      0.56 &                             263.1 &         -0.98 &                           35.3 &                           33.3 &                                   29.853 &         0.03 &    -0.03 &     36$\star$ \\
          10.00 &                              11.6 &                            8.8876 &                            4.09 &     -0.04 &                             247.3 &         -0.73 &                           57.4 &                           42.0 &                                   42.599 &         0.16 &     0.13 &            18 \\
          10.14 &                               7.9 &                           21.0170 &                            3.21 &     -0.12 &                              77.4 &         -0.97 &                           59.0 &                            5.2 &                                   13.521 &         0.84 &    -0.04 &            16 \\
          10.15 &                               1.8 &                           17.8620 &                            5.16 &      0.25 &                             390.5 &         -0.91 &                           51.5 &                           46.9 &                                   42.763 &         0.05 &    -0.19 &     47$\star$ \\
          10.48 &                               5.7 &                           11.1590 &                            1.23 &      0.03 &                             285.0 &         -0.66 &                           62.0 &                           44.1 &                                   45.386 &         0.17 &     0.01 &            34 \\
          10.71 &                              30.3 &                            5.5147 &                            0.44 &      0.69 &                             151.3 &         -0.47 &                           59.6 &                           58.7 &                                   51.511 &         0.01 &    -0.08 &            32 \\
          11.36 &                              14.4 &                            9.0172 &                            2.35 &      0.25 &                              79.9 &          0.05 &                           61.0 &                           29.8 &                                   36.680 &         0.34 &    -0.07 &             5 \\
          11.96 &                               8.0 &                           11.1150 &                            5.09 &      0.47 &                             176.4 &          0.65 &                           57.4 &                           32.7 &                                   37.418 &         0.27 &     0.40 &            31 \\
          12.00 &                              53.4 &                            6.6569 &                            5.88 &     -0.94 &                              66.3 &         -0.74 &                           47.8 &                           21.0 &                                   27.128 &         0.39 &     0.16 &            15 \\
          12.07 &                              11.6 &                            9.3873 &                            0.53 &      0.42 &                             191.7 &          0.76 &                           53.6 &                           38.0 &                                   39.212 &         0.17 &    -0.15 &             9 \\
          13.77 &                             106.6 &                            4.5605 &                            4.90 &     -0.35 &                              33.9 &         -0.16 &                           39.0 &                           36.4 &                                   32.804 &         0.03 &    -0.44 &            64 \\
          13.90 &                              93.3 &                            3.7494 &                            0.19 &      0.82 &                             106.6 &         -0.84 &                           60.1 &                           51.9 &                                   48.620 &         0.07 &    0.00 &            58 \\
\bottomrule
\end{tabular}
\end{table}

\begin{table}[ht]
\label{tab:recovery}
\vspace{0.1cm}
\caption{Recovered parameters for the \ldc{} sources. 
Rows are ordered by increasing source SNR and labelled by the injection ID in the \ldc{} dataset. Sources merging within the mission lifetime (here set to 2.5 yr) are marked with stars. For each parameter (cf. Sec.~\ref{subsec:samplingparams}), we quote median and 90\% confidence intervals. In addition, we quote the area enclosed by the 90\% contour level of the sky localization posterior $\Delta\Omega$ and the number of CPU hours required to perform each run (cf. Sec.~\ref{cpuhsec}).  
}
\footnotesize
\def\arraystretch{3}
\centering
\scalebox{0.82}{
\begin{tabular}{ccccccccccccc|r}
\toprule
$\mathrm{CPUh}$ & $\tau-\tau^{\mathrm{inj}}\,\left[\mathrm{day}\right]$ & $f_0-f_0^{\mathrm{inj}}\,\left[\mathrm{nHz}\right]$ & $\Delta\Omega\,\left[\mathrm{deg}^2\right]$ & $l\,\left[\mathrm{rad}\right]$ &                      $\sin b$ & $D_L\,\left[\mathrm{Mpc}\right]$ &               $\cos \iota$ & $m_1\,\left[M_{\odot}\right]$ & $m_2\,\left[M_{\odot}\right]$ & $\mathcal{M}_c\,\left[M_{\odot}\right]$ &               $\delta\mu$ &                 $\beta$ & $\mathrm{ID}$ \\
\midrule \midrule
                                 ${3.7}$ &                           ${1.28^{+1.56}_{-4.11}}$ &                             ${-1.0^{+6.2}_{-6.0}}$ &                                     ${6.9}$ &        ${1.4^{+0.02}_{-0.02}}$ &  ${-0.455^{+0.026}_{-0.024}}$ &         ${77.6^{+49.1}_{-28.5}}$ &    ${0.57^{+0.38}_{-0.3}}$ &      ${37.2^{+53.0}_{-13.4}}$ &        ${14.5^{+7.2}_{-7.1}}$ &            ${19.755^{+0.003}_{-0.001}}$ &  ${0.44^{+0.41}_{-0.39}}$ &   ${0.0^{+0.7}_{-0.7}}$ &         ${1}$ \\
                                 ${5.1}$ &                          ${-0.17^{+1.03}_{-3.45}}$ &                             ${-1.7^{+6.2}_{-6.1}}$ &                                    ${51.2}$ &       ${4.85^{+0.02}_{-0.02}}$ &   ${-0.067^{+0.19}_{-0.114}}$ &       ${182.3^{+166.5}_{-52.6}}$ &  ${-0.36^{+0.21}_{-0.54}}$ &     ${87.2^{+125.2}_{-31.6}}$ &      ${33.5^{+16.8}_{-16.4}}$ &            ${46.017^{+0.009}_{-0.003}}$ &  ${0.45^{+0.41}_{-0.39}}$ &   ${0.0^{+0.6}_{-0.7}}$ &        ${14}$ \\
                                ${35.9}$ &                           ${0.07^{+0.02}_{-0.06}}$ &                           ${-6.1^{+17.0}_{-21.4}}$ &                                     ${1.4}$ &       ${5.12^{+0.01}_{-0.01}}$ &  ${-0.378^{+0.012}_{-0.013}}$ &       ${166.3^{+105.5}_{-55.0}}$ &   ${0.52^{+0.42}_{-0.28}}$ &      ${48.7^{+25.8}_{-10.1}}$ &        ${27.9^{+6.8}_{-8.4}}$ &            ${31.806^{+0.002}_{-0.001}}$ &  ${0.27^{+0.31}_{-0.22}}$ &   ${0.1^{+0.1}_{-0.1}}$ &     $20\star$ \\
                                 ${4.0}$ &                           ${1.01^{+5.38}_{-7.41}}$ &                             ${-0.2^{+6.3}_{-6.4}}$ &                                    ${14.1}$ &       ${0.48^{+0.02}_{-0.02}}$ &    ${0.349^{+0.036}_{-0.04}}$ &         ${79.4^{+39.4}_{-27.2}}$ &    ${0.65^{+0.3}_{-0.29}}$ &      ${41.9^{+59.0}_{-15.1}}$ &        ${16.2^{+8.0}_{-7.9}}$ &            ${22.158^{+0.003}_{-0.002}}$ &  ${0.44^{+0.41}_{-0.39}}$ &   ${0.0^{+0.7}_{-0.7}}$ &        ${60}$ \\
                                 ${2.5}$ &                           ${-0.4^{+9.1}_{-10.69}}$ &                             ${-0.3^{+5.4}_{-5.4}}$ &                                    ${92.0}$ &       ${3.58^{+0.03}_{-0.03}}$ &   ${0.056^{+0.136}_{-0.179}}$ &        ${189.0^{+90.9}_{-62.7}}$ &   ${0.66^{+0.29}_{-0.29}}$ &     ${99.4^{+140.1}_{-36.3}}$ &      ${38.3^{+19.4}_{-18.6}}$ &              ${52.54^{+0.01}_{-0.009}}$ &    ${0.44^{+0.4}_{-0.4}}$ &   ${0.0^{+0.7}_{-0.7}}$ &        ${28}$ \\
                                 ${3.3}$ &                           ${0.44^{+1.72}_{-4.14}}$ &                             ${-0.1^{+4.8}_{-4.9}}$ &                                     ${9.6}$ &       ${0.38^{+0.03}_{-0.03}}$ &   ${-0.676^{+0.02}_{-0.019}}$ &        ${203.8^{+91.0}_{-65.5}}$ &   ${0.68^{+0.28}_{-0.28}}$ &     ${91.7^{+116.7}_{-32.8}}$ &      ${35.6^{+17.5}_{-16.6}}$ &            ${48.658^{+0.008}_{-0.003}}$ &  ${0.44^{+0.39}_{-0.39}}$ &   ${0.0^{+0.7}_{-0.7}}$ &         ${3}$ \\
                                 ${2.7}$ &                        ${-0.76^{+12.77}_{-13.73}}$ &                             ${-0.5^{+5.2}_{-5.1}}$ &                                    ${30.9}$ &       ${1.59^{+0.03}_{-0.03}}$ &   ${-0.43^{+0.057}_{-0.053}}$ &        ${121.7^{+81.0}_{-41.9}}$ &  ${-0.52^{+0.27}_{-0.42}}$ &     ${83.6^{+116.7}_{-30.4}}$ &      ${32.3^{+16.3}_{-15.7}}$ &            ${44.245^{+0.009}_{-0.009}}$ &    ${0.44^{+0.4}_{-0.4}}$ &  ${0.0^{+0.7}_{-0.7}}$ &        ${53}$ \\
                                 ${3.2}$ &                           ${0.47^{+2.22}_{-4.34}}$ &                             ${-0.0^{+6.7}_{-6.9}}$ &                                     ${4.5}$ &       ${5.05^{+0.02}_{-0.02}}$ &  ${-0.702^{+0.014}_{-0.013}}$ &         ${78.3^{+35.3}_{-25.9}}$ &   ${0.68^{+0.28}_{-0.29}}$ &      ${44.2^{+61.9}_{-16.0}}$ &        ${17.1^{+8.6}_{-8.3}}$ &            ${23.418^{+0.003}_{-0.002}}$ &    ${0.44^{+0.4}_{-0.4}}$ &  ${0.0^{+0.7}_{-0.7}}$ &        ${26}$ \\
                                 ${7.0}$ &                           ${0.01^{+0.42}_{-1.27}}$ &                             ${-1.1^{+5.1}_{-6.0}}$ &                                     ${5.2}$ &       ${6.06^{+0.01}_{-0.01}}$ &   ${0.349^{+0.026}_{-0.027}}$ &      ${325.4^{+159.2}_{-112.5}}$ &   ${-0.63^{+0.3}_{-0.33}}$ &    ${103.5^{+124.5}_{-36.7}}$ &      ${40.3^{+19.6}_{-18.3}}$ &            ${55.072^{+0.014}_{-0.004}}$ &  ${0.44^{+0.38}_{-0.39}}$ &  ${0.0^{+0.6}_{-0.6}}$ &        ${30}$ \\
                                 ${5.0}$ &                          ${-0.19^{+0.89}_{-2.67}}$ &                             ${-3.4^{+6.4}_{-6.6}}$ &                                     ${5.6}$ &       ${5.11^{+0.02}_{-0.02}}$ &   ${-0.546^{+0.02}_{-0.019}}$ &        ${188.9^{+79.9}_{-59.4}}$ &   ${0.69^{+0.27}_{-0.28}}$ &     ${82.0^{+105.1}_{-29.1}}$ &      ${32.0^{+15.6}_{-15.0}}$ &            ${43.669^{+0.008}_{-0.003}}$ &   ${0.44^{+0.4}_{-0.39}}$ &  ${0.0^{+0.7}_{-0.6}}$ &         ${2}$ \\
                                ${19.8}$ &                           ${0.04^{+0.02}_{-0.06}}$ &                             ${-1.0^{+6.7}_{-7.9}}$ &                                     ${0.6}$ &       ${3.31^{+0.01}_{-0.01}}$ &   ${0.555^{+0.006}_{-0.006}}$ &        ${187.3^{+86.9}_{-60.8}}$ &   ${-0.65^{+0.29}_{-0.3}}$ &      ${46.4^{+23.0}_{-10.9}}$ &        ${25.8^{+7.3}_{-7.4}}$ &              ${29.852^{+0.001}_{-0.001}}$ &  ${0.29^{+0.29}_{-0.25}}$ &  ${-0.1^{+0.1}_{-0.1}}$ &     $36\star$ \\
                                 ${5.2}$ &                           ${0.09^{+0.49}_{-1.55}}$ &                              ${0.4^{+4.7}_{-4.9}}$ &                                    ${18.4}$ &       ${4.09^{+0.01}_{-0.01}}$ &  ${-0.001^{+0.092}_{-0.088}}$ &       ${242.0^{+100.0}_{-79.4}}$ &   ${-0.69^{+0.3}_{-0.27}}$ &     ${82.7^{+106.2}_{-30.8}}$ &      ${30.4^{+15.7}_{-14.1}}$ &            ${42.598^{+0.009}_{-0.003}}$ &   ${0.46^{+0.38}_{-0.4}}$ &  ${0.0^{+0.6}_{-0.7}}$ &        ${18}$ \\
                                ${14.2}$ &                           ${1.03^{+0.26}_{-0.64}}$ &                              ${6.8^{+6.0}_{-7.7}}$ &                                     ${5.5}$ &       ${3.21^{+0.01}_{-0.01}}$ &  ${-0.122^{+0.259}_{-0.033}}$ &         ${62.5^{+28.2}_{-20.7}}$ &  ${-0.67^{+0.29}_{-0.29}}$ &       ${24.8^{+25.7}_{-8.4}}$ &        ${10.1^{+4.6}_{-4.3}}$ &            ${13.519^{+0.002}_{-0.001}}$ &  ${0.42^{+0.37}_{-0.37}}$ &   ${0.1^{+0.6}_{-0.7}}$ &        ${16}$ \\
                                ${45.2}$ &                           ${0.02^{+0.05}_{-0.13}}$ &                           ${-8.0^{+66.5}_{-96.4}}$ &                                    ${55.9}$ &       ${5.17^{+0.02}_{-0.04}}$ &   ${0.286^{+0.116}_{-0.206}}$ &      ${305.5^{+215.2}_{-112.8}}$ &   ${-0.55^{+0.36}_{-0.4}}$ &      ${62.2^{+25.5}_{-11.4}}$ &       ${39.2^{+8.3}_{-10.0}}$ &            ${42.762^{+0.006}_{-0.002}}$ &  ${0.23^{+0.27}_{-0.19}}$ &  ${-0.2^{+0.5}_{-0.3}}$ &     $47\star$ \\
                                 ${8.7}$ &                            ${0.0^{+0.29}_{-0.48}}$ &                             ${-2.5^{+8.6}_{-8.7}}$ &                                    ${15.6}$ &       ${1.23^{+0.01}_{-0.01}}$ &   ${0.009^{+0.089}_{-0.099}}$ &       ${287.0^{+140.1}_{-92.5}}$ &  ${-0.64^{+0.28}_{-0.33}}$ &      ${78.6^{+75.0}_{-24.2}}$ &      ${35.7^{+14.3}_{-14.6}}$ &            ${45.385^{+0.006}_{-0.004}}$ &  ${0.38^{+0.38}_{-0.33}}$ &   ${0.1^{+0.5}_{-0.6}}$ &        ${34}$ \\
                                 ${3.0}$ &                          ${-0.01^{+1.07}_{-3.59}}$ &                             ${-0.7^{+4.6}_{-4.5}}$ &                                     ${5.8}$ &       ${0.44^{+0.02}_{-0.02}}$ &   ${0.695^{+0.015}_{-0.016}}$ &        ${181.1^{+91.1}_{-58.1}}$ &  ${-0.61^{+0.27}_{-0.34}}$ &     ${95.6^{+133.2}_{-33.4}}$ &      ${38.2^{+18.2}_{-18.6}}$ &             ${51.511^{+0.01}_{-0.003}}$ &  ${0.43^{+0.41}_{-0.38}}$ &   ${0.0^{+0.6}_{-0.7}}$ &        ${32}$ \\
                                ${10.1}$ &                           ${0.15^{+0.49}_{-1.46}}$ &                            ${-12.5^{+4.9}_{-3.6}}$ &                                     ${4.3}$ &       ${2.35^{+0.01}_{-0.01}}$ &   ${0.244^{+0.027}_{-0.029}}$ &          ${78.3^{+13.9}_{-9.8}}$ &   ${0.06^{+0.08}_{-0.08}}$ &      ${66.6^{+78.1}_{-22.4}}$ &      ${27.7^{+12.5}_{-12.5}}$ &             ${36.68^{+0.006}_{-0.002}}$ &   ${0.41^{+0.4}_{-0.37}}$ &  ${0.0^{+0.7}_{-0.6}}$ &         ${5}$ \\
                                 ${5.9}$ &                           ${0.26^{+0.32}_{-0.89}}$ &                              ${1.6^{+5.1}_{-5.7}}$ &                                     ${2.1}$ &       ${5.09^{+0.01}_{-0.01}}$ &   ${0.462^{+0.013}_{-0.014}}$ &        ${192.4^{+81.6}_{-60.7}}$ &   ${0.67^{+0.29}_{-0.29}}$ &      ${67.9^{+77.6}_{-22.6}}$ &      ${28.2^{+12.6}_{-12.6}}$ &            ${37.416^{+0.007}_{-0.003}}$ &  ${0.41^{+0.39}_{-0.36}}$ &  ${0.0^{+0.6}_{-0.6}}$ &        ${31}$ \\
                                 ${2.2}$ &                           ${0.97^{+1.89}_{-4.62}}$ &                              ${0.0^{+4.1}_{-4.2}}$ &                                     ${2.0}$ &       ${5.88^{+0.03}_{-0.03}}$ &  ${-0.936^{+0.004}_{-0.004}}$ &         ${66.4^{+23.4}_{-19.0}}$ &  ${-0.72^{+0.26}_{-0.24}}$ &      ${50.9^{+67.5}_{-18.2}}$ &        ${19.9^{+9.8}_{-9.5}}$ &            ${27.128^{+0.004}_{-0.002}}$ &   ${0.44^{+0.4}_{-0.39}}$ &  ${0.0^{+0.7}_{-0.7}}$ &        ${15}$ \\
                                 ${5.3}$ &                          ${-0.05^{+0.46}_{-1.24}}$ &                              ${0.2^{+4.0}_{-4.1}}$ &                                     ${2.7}$ &       ${0.53^{+0.01}_{-0.01}}$ &   ${0.414^{+0.015}_{-0.015}}$ &        ${182.9^{+68.1}_{-57.1}}$ &    ${0.7^{+0.26}_{-0.28}}$ &      ${73.1^{+92.3}_{-25.7}}$ &      ${28.9^{+13.9}_{-13.5}}$ &            ${39.212^{+0.007}_{-0.003}}$ &   ${0.43^{+0.4}_{-0.38}}$ &   ${0.0^{+0.6}_{-0.7}}$ &         ${9}$ \\
                                 ${4.7}$ &                         ${-0.82^{+8.24}_{-10.05}}$ &                             ${-0.1^{+4.2}_{-4.3}}$ &                                     ${9.1}$ &        ${4.9^{+0.02}_{-0.02}}$ &  ${-0.349^{+0.034}_{-0.032}}$ &           ${36.9^{+6.6}_{-4.7}}$ &  ${-0.16^{+0.08}_{-0.09}}$ &      ${62.6^{+87.3}_{-23.2}}$ &      ${23.7^{+12.3}_{-11.5}}$ &            ${32.804^{+0.005}_{-0.004}}$ &    ${0.45^{+0.4}_{-0.4}}$ &   ${0.0^{+0.7}_{-0.7}}$ &        ${64}$ \\
                                 ${2.4}$ &                           ${0.49^{+6.31}_{-8.35}}$ &                              ${0.3^{+3.6}_{-3.6}}$ &                                     ${6.6}$ &        ${0.2^{+0.03}_{-0.03}}$ &   ${0.824^{+0.012}_{-0.012}}$ &         ${99.0^{+33.5}_{-27.7}}$ &  ${-0.73^{+0.26}_{-0.24}}$ &     ${91.0^{+112.2}_{-32.3}}$ &      ${35.7^{+17.3}_{-16.4}}$ &            ${48.619^{+0.007}_{-0.005}}$ &  ${0.44^{+0.39}_{-0.39}}$ &  ${0.0^{+0.7}_{-0.7}}$ &        ${58}$ \\
\bottomrule
\end{tabular}
}
\end{table}
\end{document}